\documentclass[longauth]{aa}  

\usepackage{graphicx}
\usepackage{txfonts}
\usepackage{longtable}
\usepackage{lscape}
\usepackage{caption}
\usepackage{subcaption}
\usepackage{tabularx}  %% TEST
\usepackage{array} %% TEST
\usepackage{amsmath} %% TEST
\usepackage{booktabs} %% TEST
\newcommand{\mearth}{M$_\oplus$}
\newcommand{\rearth}{R$_{\oplus}$}
\newcommand{\msun}{M$_\odot$}
\newcommand{\rsun}{R$_{\odot}$}
\newcommand{\lsun}{L$_\odot$}
\newcommand{\rhosun}{\rho_\odot}

\def\ms{\hbox{\,m\,s$^{-1}$}}         %m.s -1
\def\m2s2{\hbox{\,m$^{2}$\,s$^{-2}$}} %m2.s -2
\def\kms{\hbox{\,km\,s$^{-1}$}}       %km.s -1
\newcommand{\gcm}{\,g\,cm$^{-3}$} % density

\begin{document}

   \title{Discovery of a multi-planetary system orbiting the aged Sun-like star HD~224018\thanks{Based on observations made with the space-based telescopes \textit{Kepler}/K2, CHEOPS, and TESS, and with the HARPS-N spectrograph mounted on the Italian {\it Telescopio Nazionale Galileo} (TNG) operated by the Fundaci\'on Galileo Galilei (FGG) of the Istituto Nazionale di Astrofisica (INAF) at the Observatorio del Roque de los Muchachos (La Palma, Canary Islands, Spain)}
     }

%   \subtitle{}

    \author{M.~Damasso\inst{1} \and
    L.~Naponiello\inst{1} \and
    A.~Anna~John\inst{2,3,4} \and
    J.~A.~Egger\inst{5} \and
    M.~Cretignier\inst{6} \and
    A.~Mortier\inst{2} \and
    A.~S.~Bonomo\inst{1} \and
    A.~Collier~Cameron\inst{3,4} \and
    X.~Dumusque\inst{7} \and
    T.~Wilson\inst{8} \and
    L.~Buchhave\inst{9}\and
    B.~Nicholson\inst{10}\and
    M.~Stalport\inst{11,12}\and
    A.~Ghedina\inst{13}\and
    D.~W.~Latham\inst{14}\and
    J.~Livingston\inst{15,16,17}\and
    L.~Malavolta\inst{18,19}\and
    A.~Sozzetti\inst{1}\and
    J.~M.~Jenkins\inst{20}\and
    G.~Mantovan\inst{18,19}\and
    A.~F.~Mart\'inez Fiorenzano\inst{13}\and
    L.~Palethorpe\inst{21,22}\and
    R.~Tronsgaard\inst{23,24}\and
    S.~Udry\inst{25}\and
    C.~A.~Watson\inst{26}  %\orcid{0000-0002-9718-3266}
          }
   \institute{INAF - Osservatorio Astrofisico di Torino, Via Osservatorio 20, I-10025 Pino Torinese, Italy\\
           	\email{mario.damasso@inaf.it}
    \and School of Physics \& Astronomy, University of Birmingham, Edgbaston, Birmingham B15 2TT, UK
    \and SUPA, School of Physics \& Astronomy, University of St Andrews, North Haugh, St Andrews, KY169SS, UK
    \and Centre for Exoplanet Science, University of St Andrews, North Haugh, St Andrews, KY169SS, UK
    \and Space Research and Planetary Sciences, Physics Institute, University of Bern, Gesellschaftsstrasse 6, 3012 Bern, Switzerland
    \and Sub-department of Astrophysics, University of Oxford, Keble Rd, Oxford OX13RH, UK
    \and Observatoire Astronomique de l’Université de Genève, Chemin Pegasi 51, 1290 Versoix, Switzerland
    \and Department of Physics, University of Warwick, Gibbet Hill Road, Coventry CV4 7AL, UK
    \and DTU Space, Technical University of Denmark, Elektrovej 328, DK-2800 Kgs. Lyngby, Denmark
    \and University of Southern Queensland, Centre for Astrophysics, West Street, Toowoomba, QLD 4350 Australia
    \and Space sciences, Technologies and Astrophysics Research (STAR) Institute, Université de Liège, Allée du 6 Août 19C, 4000 Liège, Belgium
    \and Astrobiology Research Unit, Université de Liège, Allée du 6 Août 19C, B-4000 Liège, Belgium
    \and Fundaci\'on Galileo Galilei - INAF, Rambla Jos\'e Ana Fernandez P\'erez 7, E-38712 Bre\~{n}a Baja, Tenerife, Spain
    \and Center for Astrophysics | Harvard \& Smithsonian, 60 Garden Street, Cambridge, MA 02138, USA
    \and Astrobiology Center, 2-21-1 Osawa, Mitaka, Tokyo 181-8588, Japan
    \and National Astronomical Observatory of Japan, 2-21-1 Osawa, Mitaka, Tokyo 181-8588, Japan 
    \and Department of Astronomy, The Graduate University for Advanced Studies (SOKENDAI), 2-21-1 Osawa, Mitaka, Tokyo, Japan
    \and INAF - Osservatorio Astronomico di Padova, Vicolo dell’Osservatorio 5, 35122 Padova, Italy
    \and Dipartimento di Fisica e Astronomia ``Galileo Galilei'', Universit\`a degli Studi di Padova, Vicolo dell'Osservatorio 3, 35122 Padova, Italy
    \and NASA Ames Research Center, Moffett Field, CA 94035, USA
    \and SUPA, Institute for Astronomy, University of Edinburgh, Blackford Hill, Edinburgh, EH9 3HJ, UK
    \and Centre for Exoplanet Science, University of Edinburgh, Edinburgh, EH9 3HJ, UK
    \and DTU Space, National Space Institute, Technical University of Denmark, Elektrovej 328, DK-2800 Kgs. Lyngby, Denmark
    \and Affiliate member, Cavendish Laboratory, University of Cambridge, J J Thomson Avenue, Cambridge, CB3 0HE, United Kingdom
    \and Astronomy Department of the University of Geneva, Chemin Pegasi 51, CH-1290 Versoix, Switzerland 
    \and Astrophysics Research Centre, School of Mathematics and Physics, Queen’s University Belfast, Belfast, BT7 1NN, UK
    }

   \date{}

% \abstract{}{}{}{}{} 
% 5 {} token are mandatory
 
  \abstract
   {Exoplanetary systems show a large diversity of architectures and planet types. Among the increasing number of exo-demographics studies, those exploring correlations between the presence of close-in small planets and cold Jupiters are the object of particular attention. }
   {In 2016, \textit{Kepler}/K2 detected a system of two sub-Neptunes transiting the star HD\,224018, one of them showing a mono-transit event. In 2017, we began a spectroscopic follow-up with HARPS-N to measure the dynamical masses of the planets using radial velocities, and collected additional transit observations using CHEOPS.}
   {We measured the fundamental physical parameters of the host star, which is an ``old Sun'' analogue. We analysed radial velocities and photometric time series, also including data by TESS, to provide precise ephemerides, radii, masses, and bulk densities of the two planets, and possibly modeling their internal structure and composition. }
   % results
   {The system turned out to be more crowded than shown by \textit{Kepler}/K2. Radial velocities revealed the presence of two additional bodies: a candidate cold companion on an eccentric orbit with a minimum mass nearly half that of Jupiter (eccentricity $0.60^{+0.07}_{-0.08}$; semi-major axis 8.6$^{+1.5}_{-1.6}$~au), and an innermost super-Earth (orbital period 10.6413$\pm$0.0028 d; mass 4.1$\pm$0.8 \mearth) for which we discovered previously undetected transit events in \textit{Kepler}/K2 photometry. TESS data revealed a second transit of one of the two companions originally observed by \textit{Kepler}/K2. This allowed us to constrain its orbital period to a grid of values, the most likely being $\sim$138 days, which would imply a mass less than 9 \mearth\, at a 3$\sigma$ significance level. Given the level of precision of our measurements, we were able to constrain the internal structure and composition of the second-most distant planet from the host star, a warm sub-Neptune with a bulk density of 3.9$\pm$0.5 \gcm.}
   {HD\,224018 hosts three close-in transiting planets in the super-Earth-to-sub-Neptune regime, and a candidate cold and eccentric massive companion. Additional follow-up is needed to better characterise the physical properties of the planets and their architecture, and to study the evolutionary history of the system.}


    \titlerunning{A multi-planetary system orbiting HD\,224018}
    \authorrunning{Damasso et al.}

   \maketitle
%
%-------------------------------------------------------------------

\section{Introduction} \label{sec:intro}

Exoplanetary systems exhibit a vast diversity of architectures \citep{Howe2025}, varying in number, size, mass, and orbital configuration of the planets. 
Among this diversity, we find tightly packed systems with small planets in inner coplanar or mutually inclined orbits; systems with giant planets in wide, and sometimes highly eccentric orbits; and systems with both close-in small planets and outer gas giants. 
Various architectures arise from the different initial conditions in protoplanetary disks, as well as from the stochastic processes inherent to planetary formation and evolution (e.g., \citealt{Mordasini2012,Wu2019,Wang2022}). %Raymond2008
Whether the populations of close-in small planets (with mass $1$\,\mearth $< m_{\rm p} < 20$\,\mearth\ and semi-major axis $a \lesssim 0.4$~au) and cold Jupiters (i.e., outer gas giants with $m_{\rm p} =0.3-13\, \rm M_{Jup}$ and semi-major axis $a =1-10$~au) are related and to what extent, and whether any possible correlation depends on stellar parameters, such as metallicity and mass, is currently a subject of lively debate (see, e.g., \citealt{Bryan2019, Bonomo2023, BryanLee2024, BryanLee2025, VanZandtPetigura2024, Bonomo2025A&A...700A.126B}).

Among multi-planet systems with small planets in short-period orbits, those in which planets have distinct properties are especially intriguing, such as equilibrium temperatures and/or radii, in particular below and above the ``radius valley'', a marked deficit in the number of planets with sizes between about 1.5 and 2 $\rm R_{\oplus}$ in the radius distribution of small exoplanets \citep[e.g.][]{Fulton2017}.  
This radius gap reflects a dichotomy in planetary composition, in that it separates smaller Earth-like planets and super-Earths from larger sub-Neptunes, which are possibly ice-rich planets and/or planets surrounded by a substantial gas envelope. It may be explained invoking different formation processes, including orbital migration mechanisms \citep[e.g.][]{venturini2020A&A...643L...1V}, of super-Earths and sub-Neptunes (for instance, inside and outside the water snowline for rocky super-Earths and ice-rich sub-Neptunes, respectively; see e.g., \citealt{Parc2024,Shibata2025}) and/or post-formation processes, such as  atmospheric photo-evaporation \citep[e.g.][]{owen2017ApJ...847...29O,VanEylen2018}, core-powered mass loss (e.g., \citealt{Ginzburg2018,gupta2019MNRAS.487...24G}), and erosion caused by giant impacts (e.g., \citealt{Reinhardt2022}). 
The above-mentioned formation and evolutionary mechanisms are expected to produce characteristic trends in the location of the radius valley as a function of orbital period or stellar irradiation level (e.g., \citealt{Lopez2018, Martinez2019,vaneyl10.1093/mnras/stab2143}). To investigate these trends, it is crucial to determine accurate and precise bulk densities of small planets with relatively long orbital periods ($P \gtrsim 30$~d) corresponding to lower insolation levels. However, the number of well-studied planets with $P \gtrsim 30$~d remains relatively scarce compared to those with shorter periods. This is due to transit detection and follow-up surveys, such as TESS and CHEOPS, being biased toward short-period planets, and to the difficulties in detecting the Doppler radial velocity (RV) signals induced on their host star, often hampered or mimicked by those due to stellar activity.

The Sun-like star HD\,224018 (EPIC\,246214735), located at 106 pc, was initially identified as an interesting target for RV follow-up as four transit-like features were detected in the light curve collected by the space-based telescope \textit{Kepler}/K2 (hereafter K2). Three transits are produced by a sub-Neptune with a period $P$$\sim$37~d, while the fourth signal is a mono-transit of another sub-Neptune-sized companion on a wider orbit. We observed HD\,224018 with the high-resolution spectrograph HARPS-N for seven years with the aim of measuring the masses of the two companions, and concurrently better constraining the orbit of the outermost one. In parallel with our RV campaign, we collected additional photometric transits of the innermost sub-Neptune with the CHaracterising ExOPlanets Satellite \citep[CHEOPS;][]{Benz2021ExA....51..109B}. HD\,224018 was also observed by the Transiting Exoplanet Survey Satellite \citep[TESS;][]{ricker2015JATIS...1a4003R}, increasing the sample of the detected transits.    
In this study we present the results from the RV and photometric follow-up. Our data allowed us to detect two additional companions in the system: a short-period transiting Earth-size planet, and a candidate cold Jupiter on a wide and highly eccentric orbit. Our findings make HD\,224018 an interesting target for exoplanet population studies, in that it would enrich the demographics of multi-planetary systems with an eccentric and massive outer planet encircling a compact group of small-size companions.

The paper is structured as follows. In Sect. \ref{sec:datadescription} we describe the datasets used in this study. We report the fundamental physical parameters of the host star in Sect. \ref{sec:stellarparam}. In Sect. \ref{sec:rvfreqanalysis} we show the results of a frequency analysis of the RVs, and in Sect. \ref{sec:photoanalysis} we present the outcome of the analysis of transit photometry. In Sect. \ref{sec:spectrodataanalysis} we discuss the analysis of spectroscopic stellar activity diagnostics and of the RV systematics. Based on that, we define the RV dataset used for a joint spectroscopic-photometric modeling, which is described in Sect. \ref{sec:photorvanalysis}. In Sect. \ref{sec:internalstructure} we discuss the possible internal structure and composition of the only planet for which we were able to precisely measure the mass and radius. We conclude with a summary and discussion on future perspectives in Sect. \ref{sec:conclusions}.

%--------------------------------------------------------------------
\section{Observations and data reduction} \label{sec:datadescription}

\subsection{Photometry} \label{sec:photodata}

\subsubsection{K2} \label{sec:k2data}
\begin{figure*}
    \centering
    \includegraphics[width=0.8\textwidth]{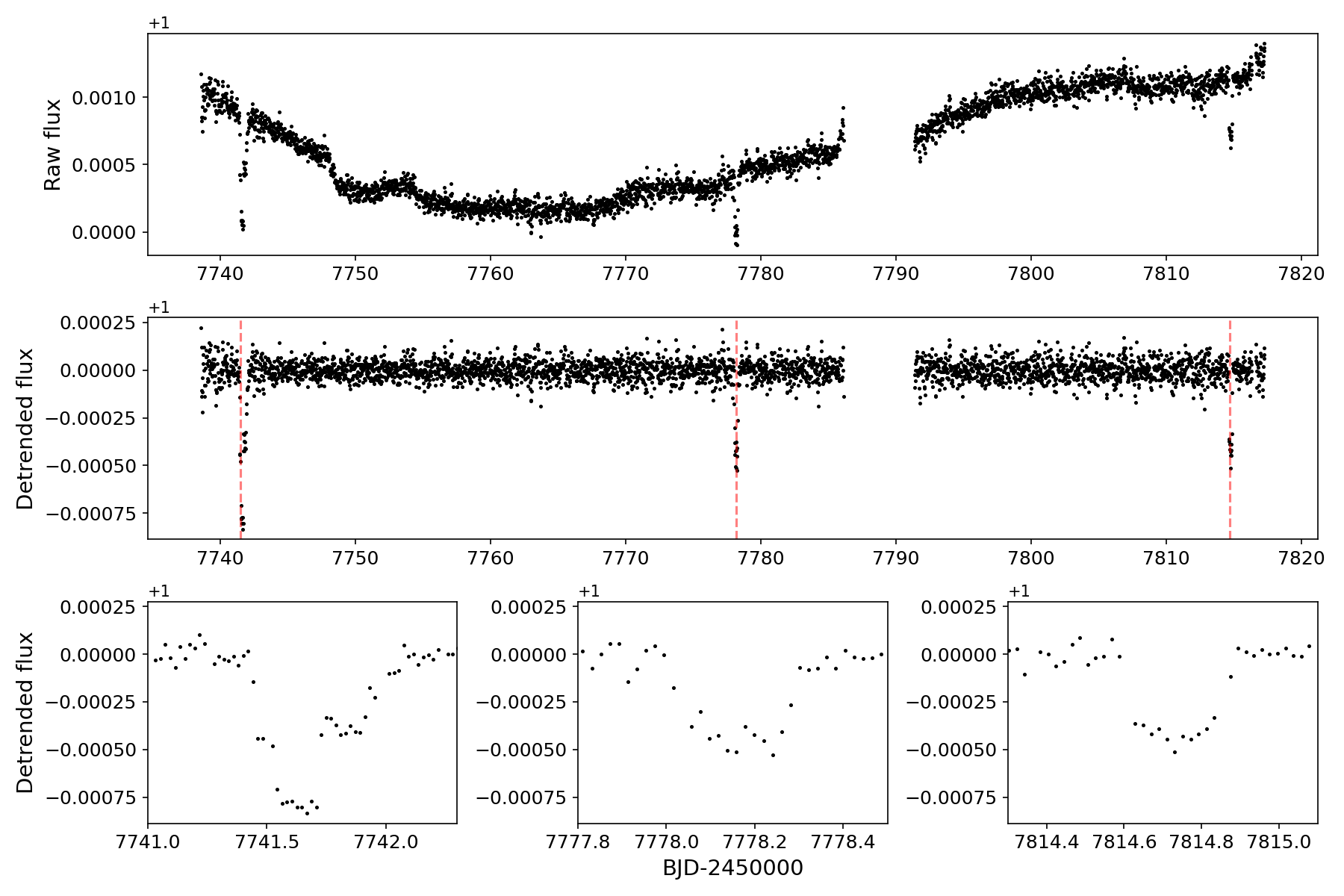}\\
    \caption{K2 light curve of HD\,224018. \textit{Upper panel.} Undetrended photometric time series. \textit{Middle panel.} A first version of the detrended and flattened light curve, as provided in the public archive (\citealt{2014PASP..126..948V}; the curve analysed in this work is shown in Fig. \ref{fig:wotandetrending}). Naked-eye visible transit signals are marked with red dashed lines. \textit{Lower panels.} Zoomed-in views showing details of the transit-like signals. The morphology of the first one looks more complex as it is the superposition of two transits. }
    \label{fig:k2lightcurves}
\end{figure*}

HD\,224018 was observed by K2 during campaign 12 from 15 December 2016 to 4 March 2017. We analysed the 30-min cadence light curve\footnote{Available at \url{https://lweb.cfa.harvard.edu/~avanderb/k2c13/ep210797580.html}} that was extracted following the methodology described by \cite{2014PASP..126..948V}. The modeling of all the transit signals detected in the photometric time series was performed on the light curve which was detrended\footnote{Hereafter we use the term ``detrended'' to refer to a light curve which has been flattened and normalised after removing temporal trends.} using the publicly available \texttt{python} package \texttt{w{\={o}}tan}\footnote{\url{https://github.com/hippke/wotan}} \citep{wotan2019AJ....158..143H}, using the built-in \texttt{cosine} function (i.e. a sum of sines and cosines, with an iterative clipping of 2$\sigma$ outliers until convergence). The final detrended light curve was obtained after masking all the identified transits (see Sect. \ref{sec:photoanalysis} for details). Fig. \ref{fig:k2lightcurves} shows the undetrended K2 light curve, a preliminary version of the detrended photometric time series, and a close-up view of the three naked-eye visible transit features. These are the signals that triggered our spectroscopic follow-up which started a few months after the detection by K2. At a first look, three of these events appear to be produced by one companion with a periodicity of $\sim$37 days, but the first one is actually deeper and due to the superposition of at least two distinct transits of a similar depth. That second event corresponds to the mono-transit of a second planet-sized companion.  

\subsubsection{CHEOPS} \label{sec:cheops}
Due to its photometric precision, the CHEOPS space telescope has been successfully used in the discovery \citep[e.g.][]{Bonfanti2021,Wilson2022,Osborn2023} and timing \citep[e.g.][]{Borsato2021,Nascimbeni2023,borsato2024A&A...689A..52B} of small exoplanets. To refine the ephemerides of the planet that transits the host every $\sim$37 days after a gap of 4.5 years from K2 observations, we covered two consecutive transit events with CHEOPS on 2-3 September 2021 and 9-10 October 2021 as part of the CH\_PR220007 program. Our visits have an observational efficiency of 90.2\% and 75.7\%, respectively, and a cadence of 60 sec resulting in 1645 total frames.

We processed both visits with the latest version of the CHEOPS Data Reduction Pipeline (DRP v14, \citealt{Hoyer2020}) that performs calibrations (e.g. dark current, flat field correction, event flagging) and corrections (background flux, cosmic rays, and smearing from nearby objects) on the raw data. The CHEOPS DRP performs aperture photometry on the images using 25 circular apertures, with radii ranging from 15 to 40 pixels. In this study, the light curves that have the lowest RMS scatter are those produced with the 26 and 23 pixel apertures, respectively. We utilised the \texttt{PYCHEOPS} package\footnote{https://github.com/pmaxted/pycheops} \citep{Maxted2022} to retrieve basis vectors corresponding to instrumental metrics (e.g. target x- and y-centroid position, background value, CHEOPS spacecraft roll angle, etc.). We used the \texttt{Juliet} python package \citep{Espinoza2019} to detrend the light curve, by simultaneously fitting the transit model alongside a linear noise model constructed from the CHEOPS basis vectors.

%UTC start:  2021-09-02T19:15:09
%UTC end:    2021-09-03T11:45:36
%Visit duration: 59428 s
%Exposure time: 1 x 60.0 s
%Number of non-flagged data points: 894
%Efficiency (non-flagged data): 90.2 %

%UTC start:  2021-10-09T09:48:09
%UTC end:    2021-10-10T02:18:36
%Visit duration: 59428 s
%Exposure time: 1 x 60.0 s
%Number of non-flagged data points: 751
%Efficiency (non-flagged data): 75.7 %

\subsubsection{TESS} \label{sec:tess}
HD\,224018 was observed by TESS in sectors 42 (20 August--16 September 2021) and 70 (20 September--15 October 2023). For our analysis we used the Simple Aperture Photometry \citep[SAP;][]{twicken:PA2010SPIE,morris:PA2020KDPH} light curve, provided by the TESS Science Processing Operations Center \citep[SPOC;][]{jenkins2016SPIE.9913E..3EJ,TESSSPOC2020} pipeline, with a cadence of 120 and 200 sec for the two sectors, respectively. The SAP light curve was not corrected for dilution because of a negligible contamination level within the photometric aperture. 
The quality of the data is not comparable to that of K2 and CHEOPS. Nonetheless, we used TESS data to improve the precision on the ephemerides of the $\sim$37~d transiting planet. Sector 70 turned out to be important for the presence of a second transit of the other companion identified in the K2 light curve.  

\subsection{Spectroscopy} \label{sec:spectrodata}
We observed HD\,224018 with the HARPS-N spectrograph \citep{Cosentino2012} from 6 August 2017 to 10 October 2023, collecting 206 spectra that were reduced with the standard Data Reduction Software (DRS) pipeline (version 3.0.1), and accessed through the Data Analysis Center for Exoplanets (DACE)\footnote{\url{https://dace.unige.ch/}} Web platform. The median and rms of the spectra signal-to-noise ratio (S/N) are $\sim$75 and 10, respectively, as measured at a reference wavelength of 5500\AA.  
The target was placed on fibre A of the spectrograph, while fibre B was simultaneously illuminated with a Fabry-Pérot etalon used as a calibration source. 
The RVs and the activity diagnostics full-width at half maximum (FWHM), bisector span (BIS), and contrast have been calculated from the cross-correlation function (CCF; \citealt{baranne1996A&AS..119..373B}) using a template mask for a star of spectral type G8, and have a median internal error $\sigma_{\rm RV}$ of 1.51 \ms and median systemic RV of -62870.84 \ms. The list of RVs and activity indexes, including the S-index measured from the CaII H and K lines and calibrated on the Mount Wilson scale \citep{wilson1968ApJ...153..221W}, is provided through the Strasbourg astronomical Data Center (CDS).

\subsection{High resolution imaging}
On the night of 14 August 2017, HD\,224018 was observed with the NESSI speckle imager \citep{Scott2019}, mounted on the 3.5\,m WIYN telescope at Kitt Peak. NESSI acquires data in two bands centered at 562\,nm and 832\,nm simultaneously using high-speed electron-multiplying CCDs (EMCCDs). We collected and reduced the data following the procedures described in \citet{Howell2011}. The resulting reconstructed image achieved a contrast of $\Delta\mathrm{mag}$$\sim$5.7 at a separation of 1\arcsec in the 832\,nm band (see Fig.~\ref{fig:wiyn}), and no secondary sources were detected.

%--------------------------------------------------------------------

\section{Stellar fundamental parameters} \label{sec:stellarparam}

\begin{table}
  \caption{Stellar parameters of HD\,224018 (EPIC\,246214735; TIC\,248608315)}
         \label{Tab:starparam}
         \tiny
         \centering
   \begin{tabular}{l l l}
            \hline
            \noalign{\smallskip}
            Parameter  &  Value & Ref. \\
            \noalign{\smallskip}
            \hline
            \noalign{\smallskip}
            RA [ICRS J2000, deg] & 358.63949 & (1) \\
            DEC [ICRS 2000, deg] & -04.72330 & (1) \\
            Parallax [mas] & 9.3912$\pm$0.0214 & (1) \\
            Proper motion [mas/yr] & -112.585, -8.324 & (1) \\
            RV [km/s] & $-62.77\pm0.28$ & (1) \\
            U [km/s] & $51.4\pm0.1$ & (2) \\
            V [km/s] & $-6.3\pm0.1$ & (2) \\
            W [km/s] & $67.1\pm0.3$ & (2) \\
%            B & 10.33$\pm$0.04 & (2) \\
%            V & 9.72$\pm$0.03 & (2) \\
            $B_{\rm T}$ & 10.459$\pm$0.043 & (3) \\
            $V_{\rm T}$ & 9.785$\pm$0.034 & (3) \\
            $B$ & 10.314$\pm$0.022 & (4) \\
            $V$ & 9.675$\pm$0.090 & (4) \\
            $g'$ & 9.980$\pm$0.078 & (4) \\
            $r'$ & 9.510$\pm$0.090 & (4) \\  $i'$ & 9.378$\pm$0.025 & (4) \\ 
            $J$ & 8.496$\pm$0.024 & (5) \\
            $H$ & 8.189$\pm$0.033 & (5) \\
            $K_{\rm s}$ & 8.145$\pm$0.027 & (5) \\
            $G$ & 9.5288$\pm$0.0028 & (1) \\
            $W1$ & 8.073$\pm$0.023 & (6) \\
            $W2$ & 8.111$\pm$0.020 & (6) \\
            $W3$ & 8.095$\pm$0.023 & (6) \\
            Effective temperature, $T_{\rm eff}$ [K] & $5784\pm60$  & (2)  \\
            Surface gravity, $\log g_\mathrm{spec}$ [$\log_{\rm 10}$cgs] & $4.30\pm0.10$ & (2)  \\
            Microturbulence, $\xi$ [\kms] & $1.15\pm0.05$ & (2)  \\
            $v~\sin i_\star$ [\kms] & $2.1\pm0.5$ & (2)  \\
            $\rm [m/H]$ & $-0.03\pm0.08$ & (2)  \\
            $\rm [Fe/H]$ & 0.05$\pm$0.05 & (2) \\
            $\rm [Mg/H]$ & 0.04$\pm$0.05 & (2) \\
            $\rm [Si/H]$ & 0.08$\pm$0.05 & (2) \\[1ex]
            Mass, $M_\star$ [\msun] & 1.013$^{+0.069}_{-0.061}$ & (2) \\[1ex]
           % & 0.999$^{+0.029}_{-0.021}$ & (6) \\[1ex]
            Radius, $R_\star$ [\rsun] & 1.147$\pm$0.028 & (2) \\[1ex]
           % & 1.156$\pm$0.008 & (6) \\[1ex]
            Density, $\rho$ [$\rhosun$] & 0.67$\pm0.07$  &  (2) \\[1ex]
          %  & 0.647$^{+0.028}_{-0.022}$  &  (6) \\[1ex]
            Luminosity, $L$ [\lsun] & $1.33^{+0.04}_{-0.05}$ & (2)  \\[1ex]
         %    $1.36^{+0.05}_{-0.04}$ & (6)  \\[1ex]
             Surface gravity, $\log g_\mathrm{iso}$ [$\log_{\rm 10}$cgs] & $4.32^{+0.04}_{-0.03}$ & (2)  \\[1ex]
        %   & $4.31^{+0.02}_{-0.01}$ & (6)  \\[1ex]
            Age [Gyr] & $7.0^{+3.4}_{-3.2}$ & (2)  \\[1ex]
       %      & $7.6^{+0.9}_{-1.2}$ & (5)  \\[1ex]
            \noalign{\smallskip}
            \hline
     \end{tabular}  
     \tablefoot{(1) from Gaia EDR3 \citep{gaia2016,gaia2021}; (2) This work;
     (3) Tycho-2 \citep{Hog2000}; (4) APASS \citep{Henden2016}; (5) 2MASS \citep{cutri2003}; (6) WISE \citep{cutri2013}.}
\end{table}

HD\,224018 is a G dwarf star located $106.1\pm0.2$\,pc away \citep{Bailer-Jones2021}. 
We derived the atmospheric stellar parameters by analysing the HARPS-N spectra: effective temperature ($T_{\mathrm{eff}}$), surface gravity ($\log g$), metallicity ([Fe/H]), microturbulence ($\xi$), and projected rotational velocity ($v~\sin i_\star$). The former three parameters were derived via two methods, and we adopt their inverse-variance weighted average as our final parameters, while the last two parameters are method-specific. First, we applied the curve-of-growth method \texttt{ARES+MOOG}. In this method, we measure the equivalent widths of neutral and ionised iron lines from the co-added spectrum. Assuming local thermodynamic equilibrium and employing the Kurucz Atlas 9 plane parallel model atmospheres \citep{Kurucz1993}, we fit for the atmospheric parameters using radiative transfer modeling. The details of the method are described in \citet{Sousa2014}. To ensure accuracy, we corrected the surface gravity following \citet{Mortier2014}. Systematic errors were added in quadrature following \citet{Sousa2011}. We find that $T_{\mathrm{eff}}=5813\pm69$\,K, $\log g=4.33\pm0.11$\,dex, and [Fe/H]$=0.05\pm0.05$. This method additionally measures the microturbulence $\xi$. Following a second method, we used \texttt{SPC} \citep[Spectral Parameter Classification - ][]{Buchhave2012, Buchhave2014}, a technique based on spectral synthesis. Each individual spectrum is analysed, and the final results are obtained by taking a weighted average of the individual spectra using the S/N values for the weights. To ensure accuracy in the surface gravity measurement, YY isochrones \citep{Yi2001} are used as an extra constraint. We find that $T_{\mathrm{eff}}=5755\pm50$\,K, $\log g=4.28\pm0.10$\,dex, and [m/H]$=-0.03\pm0.08$. This method additionally measures the projected rotational velocity.

Individual chemical abundances are useful for interior modeling of exoplanets. For these purposes, we measured the magnesium and silicon abundances in addition to the already derived iron abundance. The abundances were calculated using \texttt{ARES+MOOG} (see \citealt{Mortier2013} for details). We find that [Mg/H] = 0.04$\pm$0.05 and [Si/H] = 0.08$\pm$0.05. This is consistent with the solar-type nature of this star.

To measure the stellar radius, mass, and age, we use the effective temperature and metallicity derived from the spectra. We perform an analysis based on isochrones and evolutionary tracks, using both the Dartmouth and MIST stellar models \citep{Dotter2008,Dotter2016}. Next to the spectroscopic input, we used photometric apparent magnitudes in 8 bands [Johnson B and V, 2MASS and WISE], and the Gaia DR3 parallax \citep{Lindegren2021,Vallenari2023}. The analysis was performed four separate times, allowing for each combination of the individual spectroscopic results and the different stellar models. More details can be found in \citet{Mortier2020}. By adopting as final parameters and errors the median and 16th-84th percentiles of the combined posterior distributions, we found $M_\star=0.999^{+0.029}_{-0.021}~\rm M_\odot$, $R_\star=1.156\pm0.008~\rm R_\odot$, and age $t=7.6^{+0.9}_{-1.2}$~Gyr. 

We also modelled the MIST evolutionary tracks along with the stellar Spectral Energy Distribution (SED) in a differential evolution Markov chain Monte Carlo Bayesian framework through the {\tt EXOFASTv2} tool \citep{2017ascl.soft10003E, Eastman2019}; see \citet{Naponiello2025} for more details. To this end, we imposed Gaussian priors on the Gaia DR3 parallax as well as on the $T_{\rm eff}$ and [Fe/H] from our spectroscopically derived parameters. To sample the SED, we used the Tycho-2 $B_{\rm T}$ and $V_{\rm T}$, APASS Johnson $B$ and $V$, and Sloan $g'$, $r'$, $i'$ magnitudes, the 2MASS near-infrared $J$, $H$, and $K_{\rm s}$ magnitudes, and the WISE $W1$, $W2$, and $W3$ infrared magnitudes (Table~\ref{Tab:starparam}). The stellar SED and its best fit are shown in Fig.~\ref{fig:stellarSED}. From the medians and 16th-84th percentiles of the posteriors we derived  $M_\star=1.013^{+0.069}_{-0.061}~\rm M_\odot$, $R_\star=1.147 \pm 0.028~\rm R_\odot$, and age $t=7.0^{+3.4}_{-3.2}$~Gyr, which fully agree with the previous solution. Given the larger, and thus more conservative, uncertainties, we chose these as our reference stellar parameters in this work. The derived surface gravity $\log{g}=4.32^{+0.04}_{-0.03}$ is fully consistent with the spectroscopic value, but more precise.

Finally, we derived Galactic space velocities for HD\,224018 using the coordinates, proper motions, parallax and radial velocity from Gaia DR3. We followed the formulation by \citet{Johnson1987} to calculate the $U$, $V$, and $W$ heliocentric velocity components and its errors. The velocities are defined in the directions of the Galactic center, Galactic rotation (using the right-hand system), and north Galactic pole, respectively. The Solar motion was not subtracted. From these velocities we assign membership probabilities using a Monte Carlo approach and following the \citet{Reddy2006} formalism. Kinematically, this star is found to have a probability of $88.4\pm0.5\%$ to be a thick disc star and $11.1\pm0.5\%$ to be a thin disc star. The age and chemical composition of this star point more to a thin disc star, but could still be consistent with a thick disc star.
All our derived stellar parameters are summarised in Table \ref{Tab:starparam}. 

%From these fundamental parameters, we can now calculate a new value for the surface gravity. It is 

%ARESMOOG
%Name	Teff	erteff	logg	erlogg	vt	ervt	feh	erfeh
%----	----	------	----	------	--	----	---	-----
%EP246214735_S1D_sum	5813.0	69.0	4.33	0.11	1.15	0.05	0.05	0.05

% SPC
% analyzed the 206 HARPS-N spectra and 202 of those yielded reliable parameters.
%Teff=5755 +- 50 K, log(g)=4.28 +- 0.10, [m/H]=-0.03 +- 0.08, vsini=2.1 +- 0.5 km/s

% isochrones
%MASS (solar)
%0.999431989572708 -0.02093273880635793 0.02873913662896821
%RADIUS (solar)
%1.1555390530104241 -0.008144499493453194 0.007870947748241042
%DENSITY (solar)
%0.6473620713149932 -0.022257682170013116 0.028164908306279268
%DISTANCE (pc)
%106.48489298814366 -0.2345781334159227 0.2414232398099898
%AGE (Gyr)
%7.565089717405804 -1.1919916507855568 0.9369929661488312
%METALLICITY)
%0.03538267374655872 -0.04820283123294607 0.04649577384033654
%%LUMINOSITY (solar)
%1.3585094228109793 -0.03712877835083361 0.04728771711957869
%EXTINCTION AV
%0.10561376279322522 -0.054934736187290535 0.05317950889983426
%NEW TEFF (K)
%5798.653224223588 -43.76392621022751 58.38237539986585
%NEW LOGG (CGS)
%4.311962282999171 -0.012864243069822479 0.01620767477960605

%--------------------------------------------------------------------
\section{Frequency analysis of radial velocities}  \label{sec:rvfreqanalysis}
\begin{figure}[h!]
    \centering
    \includegraphics[width=0.5\textwidth]{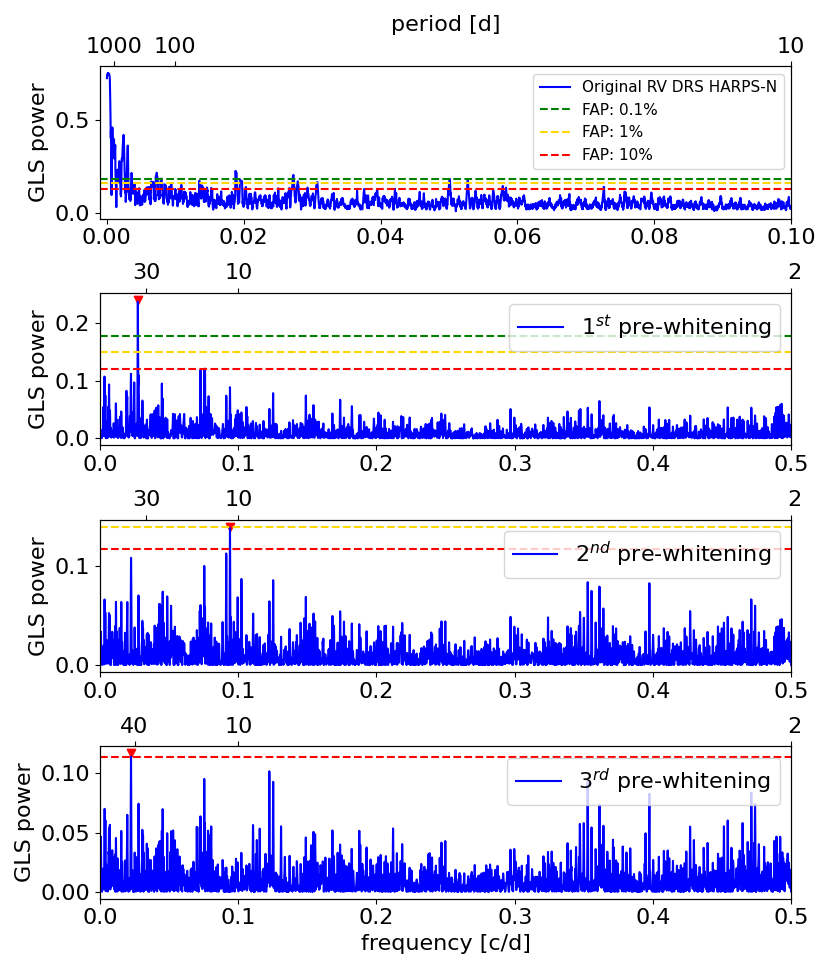}
    \caption{GLS periodograms (in frequency and period) of the HARPS-N original radial velocities (\textit{upper panel}) and residuals. These are calculated through an iterative pre-whitening. Red triangles mark the peaks with the highest power. False alarm probability levels are indicated by horizontal dashed lines.}
    \label{fig:rvprewitheperiodogram}
\end{figure}

We used the \texttt{generalised Lomb-Scargle} periodogram \citep[GLS;][]{zech2009} for a first identification of significant periodical signals in the original RV time series and pre-whitened RV residuals. Looking at the original RV time series (see the upper left panel of Fig. \ref{fig:rvplots2}), there is a clear long-term trend for which it is not possible to confirm a periodicity, because the data do not cover a sufficiently long time span. The GLS periodogram of the original dataset is shown in the first panel of Fig. \ref{fig:rvprewitheperiodogram}. We used a Keplerian with a least-squares best-fit period $P$=10998 d and eccentricity $e$=0.7 for a first pre-whitening, to search for other periodic signals. The GLS periodograms from the second to the fourth pre-whitened dataset are shown in Fig. \ref{fig:rvprewitheperiodogram}. We adopted a Keplerian with zero eccentricity as a model for the iterative pre-whitening. False alarm probability (FAP) levels of the main peaks were evaluated through a bootstrap (with replacement) simulations. The analysis of the pre-whitened RV residuals reveals the existence of a significant signal at $P=36.58\pm0.07$ days (semi-amplitude $K=1.9\pm0.2$ \ms), which is a period very close to that of the transit signal detected by K2. Therefore, it is safe to conclude that it is produced by the same companion. We also found a signal at $P=10.636\pm0.007$ d with a FAP of 1.54$\%$ (semi-amplitude $K=1.1\pm0.2$ \ms), and its nature is unveiled in Sect. \ref{sec:photoanalysis}. After a third and last pre-whitening, the periodogram reveals a peak at $P=44.7\pm0.2$ d with a large FAP ($\sim10\%$). We lack additional evidence to support the hypothesis that this corresponds to the stellar rotation period or any other astrophysical signal, such as an additional planet, and given its high FAP value we will not consider it further in our analysis.

%\begin{figure}
%    \centering
%    \includegraphics[width=0.45\textwidth]{rv_hn.png}
%    \caption{\textbf{Mario: this plot could be removed. The RV time series reappears in Fig. 6. } Time series of the radial velocities measured from HARPS-N spectra.}
%    \label{fig:rvonlytimeseries}
%\end{figure}

%--------------------------------------------------------------------
\section{Photometric data analysis} \label{sec:photoanalysis}

\subsection{Improving the sensitivity to transit detection in K2 data} \label{sec:k2lcanalysis}

In Sect. \ref{sec:k2data} we showed the existence of two companions orbiting HD\,224018 thanks to the K2 observation of transits. For one of them, we detected the counterpart signal in the RV time series, while for the other companion only a mono-transit is revealed. We used the \texttt{python} package \texttt{w\={o}tan} \citep{wotan2019AJ....158..143H} to perform an optimal detrending of the K2 and TESS light curves, with the goals of preserving the actual shape of the transit profiles after removing trends from time-series data, and search for possibly undetected transits of short-period planets, in particular of a third body that could be responsible for the 10.6-day signal found in the RVs.
We found that among several of the available detrending algorithms provided by \texttt{w\={o}tan}, the \texttt{cosine} method produced the best results in terms of signal detection efficiency (SDE)\footnote{For detrending the data we used the following \texttt{w\={o}tan} keywords: window length=1.2, break tolerance=0.5, and robust=True}. After it was applied a first time to the original light curve, we used the \texttt{transit least squares} (TLS) algorithm \citep{tls_2019A&A...623A..39H} to constrain the properties of the signal due to the $\sim$36.6-d companion (which is detected with an SDE=25.9), and mask the corresponding three transits, also adopting a time interval wide enough to mask the overlapped transits. Then, we detrended the masked light curve, and running TLS on the flattened dataset led to a more significant detection of the $\sim$36.6-d signal, boosting the SDE to a value of 29.1. After masking the transits again on the better detrended light curve, we searched for additional transit-like signals with TLS. We found the main peak in the periodogram at $P$=21.28 days (SDE=8.1), i.e. at exactly twice the period of the 10.64 day signal detected in the RV data. We interpret this result as evidence that the RV signal is associated with a short-period transiting companion, and the low SDE peak revealed at $\sim$21 days is due to the low S/N of the transits, with some of them not identified by TLS. After identifying the transits of the 10.6-day companion with TLS, we masked all the transit signals in the K2 data to perform a last and more accurate detrending of the original light curve. That resulted in an increase of the SDE for the $P$=21.28 day signal (SDE=9.4). The SDE of the 10.64 d transit signal is $\sim$8. Further application of the TLS algorithm did not show other significant peaks in the periodogram.

\begin{figure}
%    \begin{minipage}[t]{.45\textwidth}
        \centering
        \includegraphics[scale=0.5]{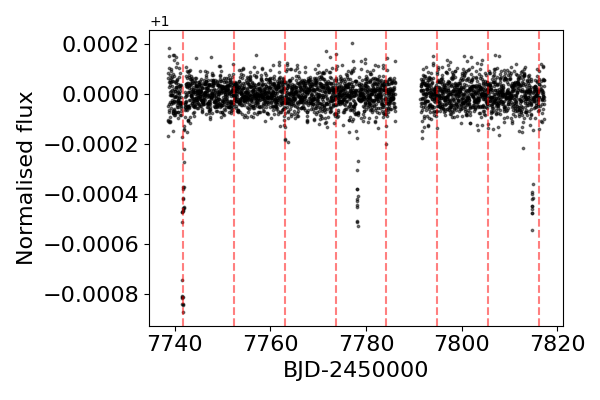}\\
%        \subcaption{}
%    \end{minipage}
%    \hfill
%    \begin{minipage}[t]{.45\textwidth}
%        \centering
        \includegraphics[scale=0.46]{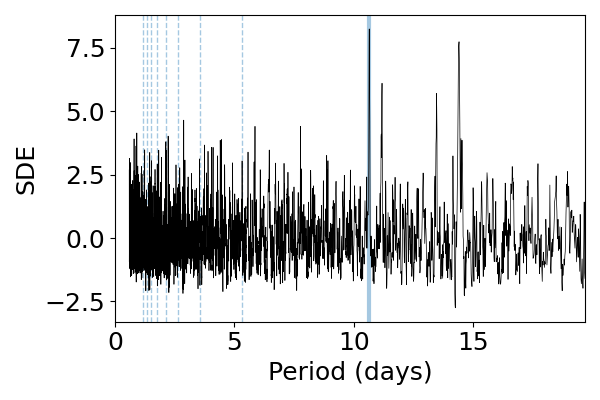}\\
%        \subcaption{}
%    \end{minipage}  
%    \begin{minipage}[t]{.45\textwidth}
%        \centering
%       \includegraphics[scale=.46]{detrended_k2_lc_with_10d_transits_nr_3.png}\\
%        \subcaption{}
%    \end{minipage}  
%    \begin{minipage}[t]{.45\textwidth}
%        \centering
%        \includegraphics[scale=0.5]{detrended_k2_lc_with_10d_transits_nr_6.png}
%        \subcaption{}
%    \end{minipage}  
    \caption{Results from the detrending of the K2 light curve with \texttt{w\={o}tan}. \textit{Upper panel}. Detrended light curve, with the epochs of the transits of HD\,224018\,b indicated by vertical dashed lines. \textit{Lower panel}. TLS periodogram after masking the 36.6 day transits and the mono-transit, and restricting the search to orbital periods shorter than 19.7 days. The main peak is found at 10.64 days. Individual transits with very low S/N transits are shown in Fig. \ref{fig:plbsingletrans}, demonstrating how a fine-tuned detrending was crucial for their detection.}
    \label{fig:wotandetrending}
\end{figure}

A summary of the results from the analysis described above is provided in Fig. \ref{fig:wotandetrending}. Our fine-tuned detrending of the K2 light curve allowed us to identify very shallow transits of the 10.64-day companion first detected in the RV, that were missed by an initial blind search (these are shown in Fig. \ref{fig:plbsingletrans}). It turns out that the first transit of the 10.64 day companion occurred while the other two companions were also transiting, so that K2 actually observed a triple transit event. Following an order based on their distance from the host star, we identify the companions with orbital periods 10.64 and 36.6 days, and the one showing a mono-transit as HD\,224018\,b, HD\,224018\,c and HD\,224018\,d, respectively. Our final measurements of radii and masses for these companions, described in Sect. \ref{sec:photorvanalysis}, justify that they are identified as planets.   

\subsection{CHEOPS and TESS light curves} \label{sec:cheopstesslcanalysis}
CHEOPS and TESS light curves were detrended with \texttt{w\={o}tan} using the same set-up and steps adopted for K2 data.
We note that the first transit of HD\,224018\,c observed by CHEOPS occurred in a gap of the TESS data in sector 42. As mentioned, TESS observations of sector 70 contain one transit, increasing to six the number of recorded transits for this planet. In Sect. \ref{sec:photorvanalysis} we will report the detection of a second transit of HD\,224018\,d in TESS data of sector 70.  

%\begin{figure}
%\begin{minipage}[t]{.5\textwidth}
%        \centering
%        \includegraphics[width=\textwidth]{cheops_lc.png}
%        \subcaption{}
%    \end{minipage}  
%    \begin{minipage}[t]{.5\textwidth}
%        \centering
%        \includegraphics[width=\textwidth]{tess_lc.png}
%        \subcaption{}
%    \end{minipage}  
%    \caption{Panel (a). Transit light curve of HD\,224018\,b observed by CHEOPS. Panel (b) Light curve of HD\,224018 observed by TESS. The transit of HD\,224018\,b is marked by a red vertical dashed line in the datset from Sector 70.  }
%    \label{fig:cheopstess}
%\end{figure}
 
%--------------------------------------------------------------------

\section{Spectroscopic data analysis} \label{sec:spectrodataanalysis}

\subsection{Activity diagnostics analysis}
\begin{figure*}[h!]
    \centering
    \includegraphics[width=0.85\textwidth]{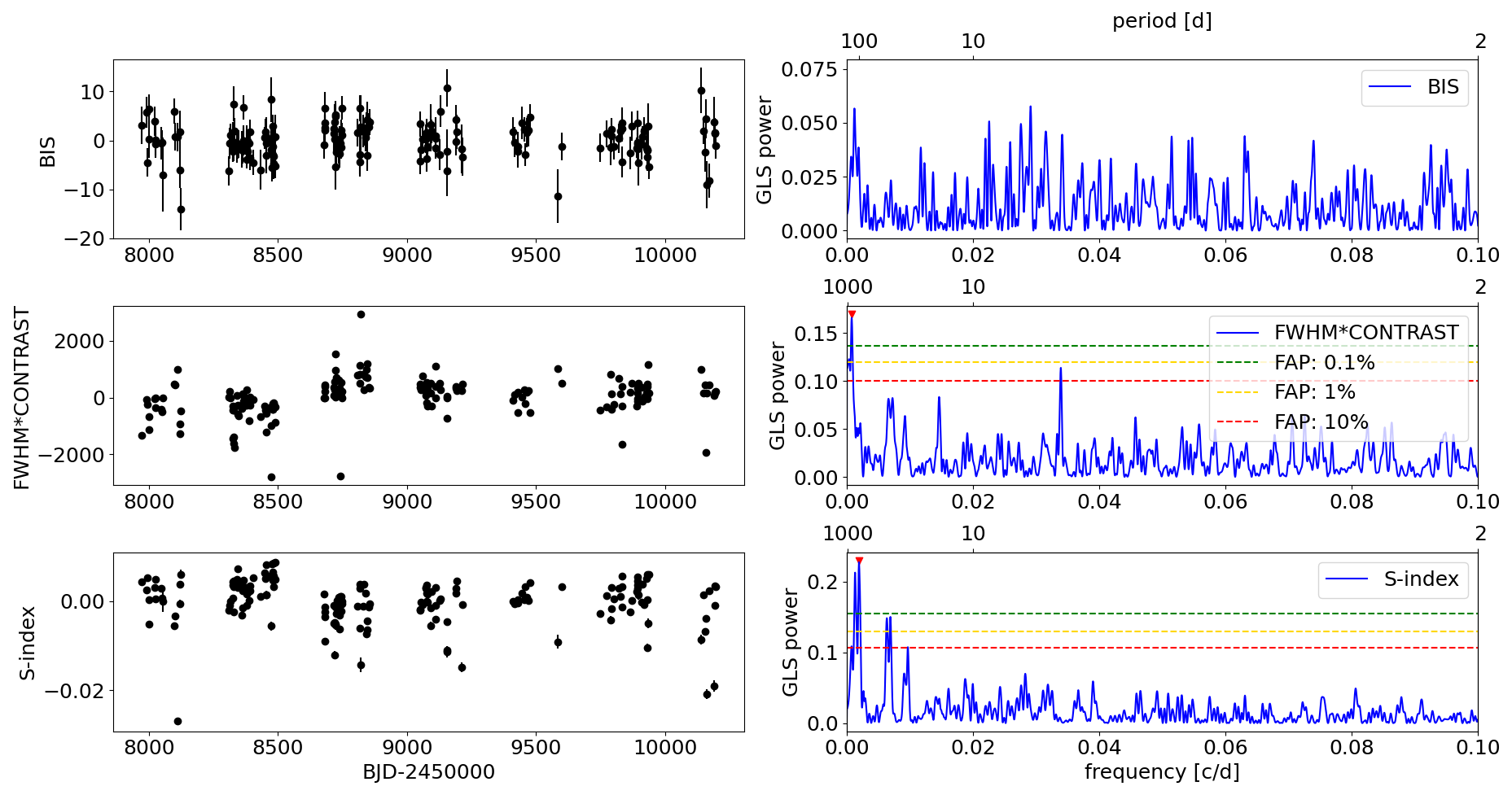}
    \caption{Time series and periodograms of the spectroscopic activity diagnostics BIS, CCF area (FWHM$\cdot$contrast), and S-index. \textit{First column.} Time series of the activity indexes. The average values have been subtracted from each dataset. \textit{Second column.} GLS periodograms. Red triangles indicate peak values. The horizontal lines correspond to levels of FAPs determined through a bootstrap analysis.}
    \label{fig:activitydiagnostics}
\end{figure*}

Figure \ref{fig:activitydiagnostics} shows the time series and the corresponding GLS periodograms of three activity indicators extracted from the HARPS-N spectra: the CCF-derived BIS and CCF area, given by the product FWHM$\cdot$contrast, and the S-index. We used the CCF area as a diagnostic, instead of the individual FWHM and contrast time series, because their significant anti-correlation is likely due to long-term changes in the instrument focus (for more details, refer to \citealt{10.1093/mnras/stz1215}). We do not find significant periodicities in the BIS time series, while the periodograms of the CCF area and S-index show significant peaks at $\sim$1255 and $\sim$515 days, respectively, with low FAPs determined through a bootstrap analysis.

\subsection{Identification of RV systematics with \texttt{YARARA} and \texttt{SCALPELS}} \label{sec:yararascalpels}
%In this section we want to  demonstrate that a long-term signal survives even after identifying the two main jumps.
%should we mention that the 36.5-d signal is significant in the RV alone, i.e. we would detect the planet without the need for a transit?

% Contribution by Michael
We processed the HARPS-N spectra time series through \texttt{YARARA}, a post-processing data-driven methodology aiming to extract as much as possible information contained in a high-resolution spectrum. The final goal of \texttt{YARARA} is to provide an independent measurement of the RV time-series, either measured by a CCF using a tailored line-selection \citep{cretignier2020A&A...633A..76C} or line-by-line RVs \citep{cretignier2023A&A...678A...2C} following the prescription of \cite{dumusque2018A&A...620A..47D}, but also providing activity indicators \citep{cretignier2024MNRAS.535.2562C}). Because of its limitation due to the data quality, we decided not to use \texttt{YARARA} to measure the RVs. Nevertheless, the analysis revealed interesting results which are reported in Appendix \ref{app:yarara}, together with more details on the method. 

\begin{figure}[h!]
    \centering
    \includegraphics[width=0.5\textwidth]{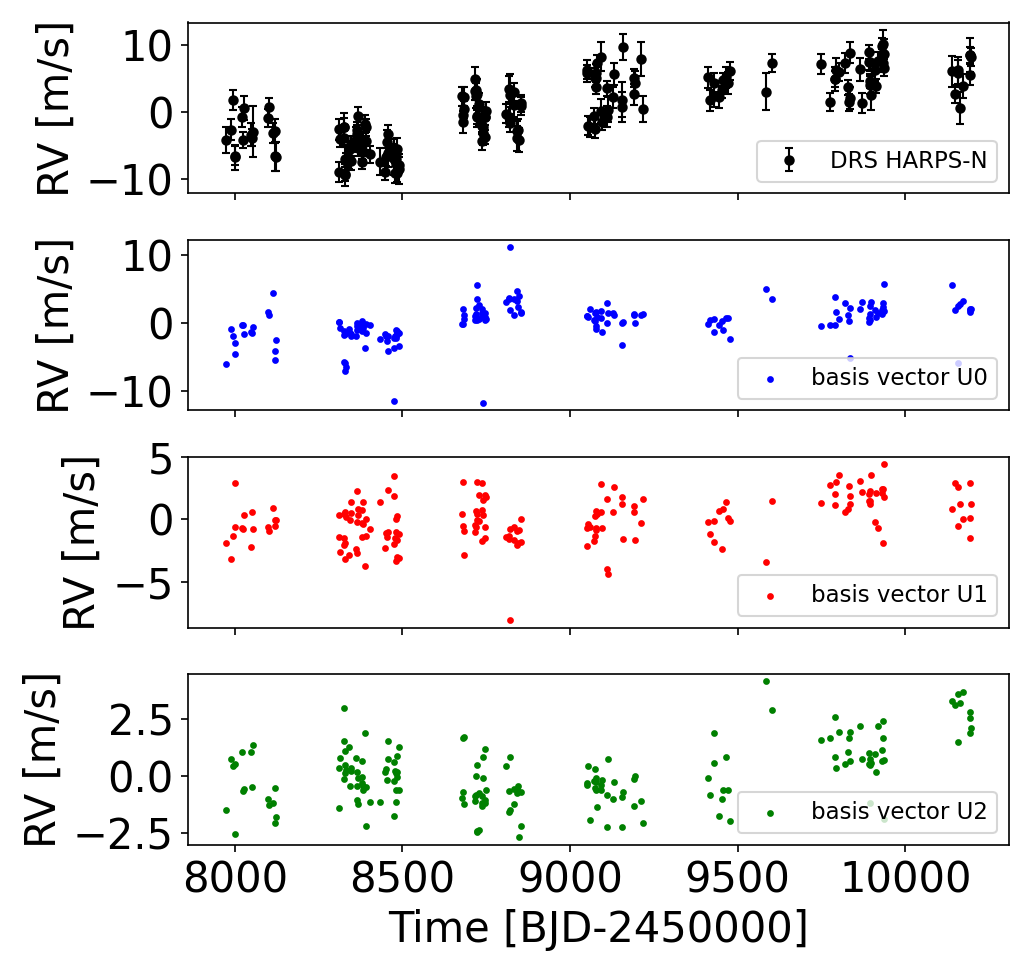}
    \caption{Radial velocity time series selected with \texttt{SCALPELS} as our definitive dataset (upper panel). Radial velocities projection onto the first three basis vectors calculated with \texttt{SCALPELS} are shown in the other panels.}
    \label{fig:rvtimeseries}
\end{figure}

%\texttt{SCALPELS}
Motivated by the need to understand the contribution of stellar activity in the RV observations, we used the \texttt{SCALPELS} method devised by \citet{ACC2021}. \texttt{SCALPELS} decorrelates for spectral line-shape changes in CCFs, induced by stellar variability, while preserving the Doppler shift component that we can use for planet search. To obtain reliable detections of planets from the \texttt{SCALPELS} `shift' component of the RVs, we used {\sc tweaks} \citep{AnnaJohn2022, AnnaJohn2023}, a workflow that integrates \texttt{SCALPELS} with the trans-dimensional nested sampling algorithm called \texttt{KIMA} \citep{Faria2016}. \texttt{SCALPELS} sets out by creating an orthogonal basis with the coefficients of the first few leading principal components of shape-induced CCF fluctuations. The raw RVs are then projected on this basis to generate a time series of shape-induced RV variations (Fig. \ref{fig:rvtimeseries}). The basis vectors (U-vectors) representing the shape components identified by \texttt{SCALPELS} are then used for stellar activity decorrelation using the \texttt{KIMA}. \texttt{KIMA} employs diffusive nested sampling \citep{brewer2014arXiv1411.3921B} to sample the posterior distributions for each of the orbital parameters by modeling the RV data with a sum of up to $N_p$ Keplerian functions. \texttt{KIMA} also perform simultaneous Gaussian Process analysis to counteract any leakage of shift-like stellar activity signal during the \texttt{SCALPELS} U-vector decorrelation. \\
To ensure a clean set of basis vectors, \texttt{SCALPELS} uses median absolute deviation (MAD) to discard anomalous observations that may generate spurious basis functions (See Section 3.2 in \citealt{ACC2021}). For HD\,224018, this resulted in the rejection of 29 spectra from the initial sample of 206, leaving 177 useful data points to use in our analysis. After the rejection of bad rows, the columns of the right singular matrix U are re-ordered in the descending order of their contribution to improve the chi-squared and the Bayesian information criterion (BIC) in a planet-free model. The number of decorrelation vectors that minimise the BIC in our case is three. The time series of RVs and the \texttt{SCALPELS} decorrelation basis vectors are shown in Fig. \ref{fig:rvtimeseries}, and provided through the CDS. 
As more planet signals are fitted, the relative significance of the basis vectors could vary because the $U$ vectors are not necessarily orthogonal to the parts of the signal that Keplerians are now fitting out. 

For HD\,224018, two basis vectors were excluded from further runs in order to avoid over-fitting, since they did not significantly contribute to the three-planet model.
Therefore, we only included $U_0$, as it may help mitigate an instrumental effect or a long-term activity cycle. Apparently, there is a long-term effect manifesting itself as a slightly asymmetric change, as it is also observed in the time series of the CCF area (cfr. with Fig. \ref{fig:activitydiagnostics}). Otherwise, the star does not display signs of strong intrinsic variability in terms of changes in the shape of the CCF. There is very little shape-driven RV variability suggesting little spot activity or photometric effect due to facular contrast. Later in this paper, we directly use the RV dataset defined with \texttt{SCALPELS} to perform the joint analysis discussed in Sect. \ref{sec:photorvanalysis}.

%Since the activity signals are predominantly shift-like, they are best fitted with a GP.
%In light of these understandings, we employed a quasi-periodic GP regression combined with a model that included up to three unknown Keplerian signals (in addition to the known transiting planets), while simultaneously performing stellar activity decorrelation against one {\sc \texttt{SCALPELS}} basis vector.
%The joint posteriors now detected a Keplerian signal at orbital period 10.64 days, corresponding to the transiting planet b, a second signal at 36.58 days of transiting planet c, and a long-term keplerian at 2829-day orbital period. The signals of planets b and c were detected with high significance with RV semi-amplitudes of 1.16 $\pm$ 0.22 and 1.82 $\pm$ 0.31 ms$^{-1}$ respectively. There is also a long-term signal present, with an orbital period of $\sim$ 2950 days, and an RV semi-amplitude of $5.9^{+0.90}_{-1.8}$ ms$^{-1}$ . The GP hyper parameter, which models the periodic time-scale of the correlated signal, converges at 44 days. Nevertheless, we lack additional evidence to back up the hypothesis that this is the star's rotation period.
%

%---------------------------------------------------------------------

\section{Photometric and spectroscopic joint analysis} \label{sec:photorvanalysis}

We measured the fundamental planetary parameters by performing a joint Monte Carlo (MC) modeling of RV and photometric light curves, using the nested sampler and Bayesian inference tool \texttt{MULTINEST V3.10} (e.g. \citealt{Feroz2019}), through the \texttt{pyMULTINEST} wrapper \citep{Buchner2014}. The MC set-up is characterised by 500 live points, a sampling efficiency of 0.5, and an evidence tolerance of 0.5. Radial velocity and transit signals were modelled using the \texttt{python} packages \texttt{radvel}\footnote{\url{https://radvel.readthedocs.io/en/latest/index.html#}} \citep{radvel2018PASP..130d4504F} and \texttt{pytransit}\footnote{\url{https://github.com/hpparvi/PyTransit}} \citep{pytransit2015MNRAS.450.3233P}, respectively. 

The analysed dataset include the 177 RVs defined through the analysis with \texttt{SCALPELS}, and the detrended K2, CHEOPS, and TESS light curves. We modeled transit signals of HD\,224018\,b only in K2 data because their S/N is too low to make them detectable in the other datasets. We also included the \texttt{SCALPELS} decorrelation basis vector $U_0$, with the linear correlation coefficient $\alpha_0$ as a free parameter. 

In summary, our RV model includes four Keplerians to fit the RV Doppler signals, and the correction of systematics with \texttt{SCALPELS}. For the photometric part, we included three transiting signals for the three innermost companions. For the eccentricities of the three transiting planets we adopted a Gaussian prior $\mathcal{N}(\mu=0,\sigma=0.1)$ based on the results of \cite{vaneylen2019AJ....157...61V}, while for the candidate companion HD\,224018\,e, instead of fitting directly the eccentricity $e_e$ and the argument of periastron $\omega_{\rm \star,\, e}$, we used the alternative parameterisation $\sqrt{e_{\rm e}}\cos\omega_{\rm \star,\:e}$ and $\sqrt{e_{\rm e}}\sin\omega_{\rm \star,\:e}$ \citep{anderson2011ApJ...726L..19A}. 
Concerning the light curves, we adopted a quadratic law for the limb darkening, and fitted the coefficients $u_{\rm 1}$ and $u_{\rm 2}$ (two sets, one in common for K2 and CHEOPS, one for TESS) using the parametrization for coefficients q$_1$ and q$_2$ given by \cite{kipping2010} (see Eq. 15 and 16 therein). We also introduced offsets $\gamma$ and constant jitters $\sigma_{\rm jit}$ as free parameters for both the spectroscopic and photometric dataset, and for each instrument. The jitter values are added in quadrature to the RV and light curve internal uncertainties. Considering the long cadence of the K2 light curve, in order to mitigate morphological distortions when fitting the transit light curves in this dataset (see \citealt{kipping2010}), we used a photometric oversampling factor of 15. 

The total number of fitted parameters of our model is equal to 40. We adopted uniform priors for all of them with the exception of the eccentricities of the three transiting companions, as mentioned above, and the stellar density $\rho_\star$, that we used in place of the $a_{\rm c}/R_{\star}$ ratio at each step of the MC sampling (e.g. \citealt{2007ApJ...664.1190S}), for which we  adopted a Gaussian prior based on the result of Sect. \ref{sec:stellarparam}. The complete list of priors and best-fit results, for free and derived parameters, is given in Table \ref{tab:resultjointfit-compact}. 

%Results
The main results of the joint RV and photometric modeling can be summarised as follows. For HD\,224018\,b, we measure a mass of 4.1$\pm$0.8 \mearth, but the radius (as well as the bulk density) cannot be determined with either accuracy or precision ($r_b$=0.89$^{+0.64\,(+1.57)}_{-0.57\,(-0.90)}$ \rearth, with 3$\sigma$ error bars given in parenthesis). Individual transits are shown in Fig. \ref{fig:plbsingletrans}. Some are not visible, such as the first and second transit, because of the very low signal S/N. No significant transit time variations are detected within the uncertainties of the retrieved transit times. HD\,224018\,c is a sub-Neptune for which we measure precise mass, radius, and bulk density: $m_c$=10.4$^{+1.3}_{-1.1}$ \mearth, $r_c$=2.42$^{+0.07}_{-0.08}$ \rearth, and $\rho_c$=4.0$^{+0.7}_{-0.6}$ \gcm. For HD\,224018\,d, our analysis reveals that the TESS data of Sector 70 likely contains a second transit. The marginal posterior of $P_d$ is shown in Fig. \ref{fig:posterior_pd}, together with the grid of possible orbital periods derived from the relation $P=\frac{1}{q}(T_{0,\,K2}-T_{0,\,TESS}$), where $q$
is an integer representing a specific harmonic of the orbital period, and $T_{0,\,K2}$ and $T_{0,\,TESS}$ are the transit midpoints of the two observed transits. From this distribution we constrain the orbital period to be $P_d$=138.0731$^{+27.6127}_{-0.0050}$ days. Fig. \ref{fig:posterior_pd} shows that $P_d=138.07$ days is the most likely orbital period among the predicted values (corresponding to the integer $q=18$), and the large asymmetry of the 1$\sigma$ error bars is determined by the second most likely period in the grid, $P_d=165.69$ days (corresponding to the integer $q=15$). The statistical preference for $P_d=138.07$ days is determined from the information contained in the RVs, from which   
we measured a semi-amplitude of the Doppler signal associated to planet d with a significance level of 2.6$\sigma$, $K_d$=0.52$^{+0.22}_{-0.20}$ \ms. This corresponds to a planetary mass $m_d$=4.2$^{+1.8}_{-1.6}$\,\mearth. The radius of this planet is comparable to that of planet c, and the resulting bulk density of HD\,224018\,d is $\rho_d$=1.7$^{+0.8}_{-0.6}$ \gcm. Nonetheless, given the low statistical significance of $K_d$, ours should be assumed as a tentative detection of the RV signal, as well as $3\sigma$ upper limits should be considered for the mass and density, 9.0 \mearth\,and 4.2 \gcm\,respectively. Indeed, whether the real orbital period of HD\,224018\,d is close to 138 days (with implications on the mass and bulk density measurements) must be confirmed with the detection of at least one more transit event.   
For the outermost candidate companion, we did not observe a full orbital cycle, thus we cannot well constrain the orbital period and semi-major axis with our current RV time series ($P_e$=9129$^{+2499}_{-2479}$ days; $a_e$=8.6$^{+1.5}_{-1.6}$ au). However, the spectroscopic orbit has a significantly high eccentricity ($e_e$=0.60$^{+0.07}_{-0.08}$), and the Doppler semi-amplitude is well constrained ($K_e$=5.7$^{+0.4}_{-0.3}$ $\ms$), corresponding to a minimum mass of nearly 0.5 M$_{\rm Jup}$. Therefore, our solution shows that HD\,224018\,e is compatible with being a cold Jupiter, although additional spectroscopic follow-up is needed to better constrain its minimum mass and orbital architecture.  
We note that the applied RV correction using the first \texttt{SCALPELS} basis vector $U_0$ is significant at a level of $\sim$2$\sigma$ ($\alpha_0$=0.13$^{+0.07}_{-0.06}$), and that the applied correction has preserved the signal of the wide-orbit planetary candidate signal, supporting that it has an astrophysical origin.   
The GLS periodogram of the RV residuals, after subtracting all the spectroscopic signals that are included in our model, confirms the presence of a statistically insignificant main peak at $\sim$45 days, as discussed in Sect. \ref{sec:rvfreqanalysis}.

\begin{figure*}
       \centering
        \includegraphics[width=0.33\textwidth]{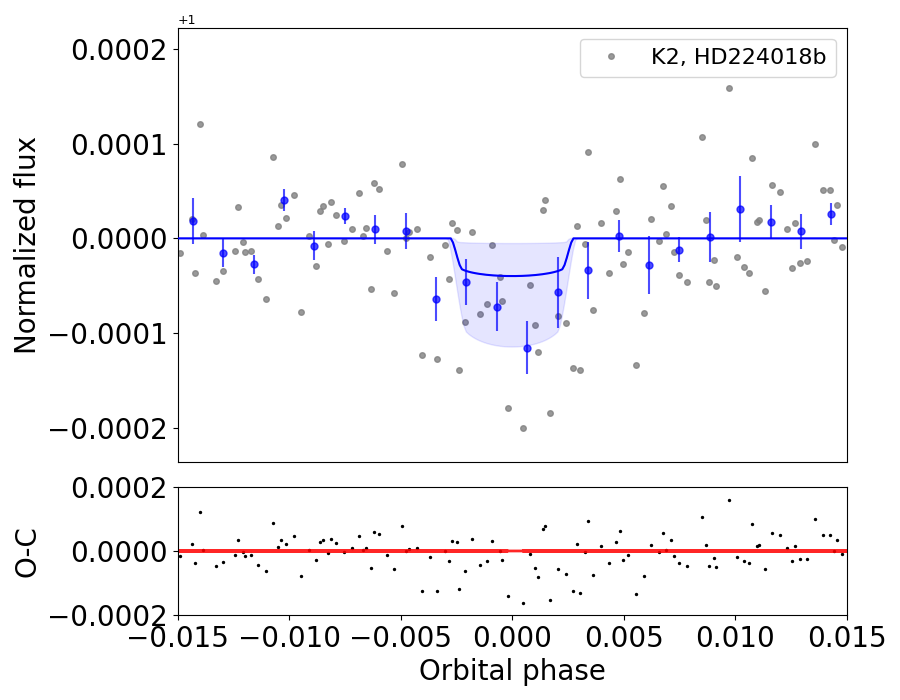} 
        \includegraphics[width=0.33\textwidth]{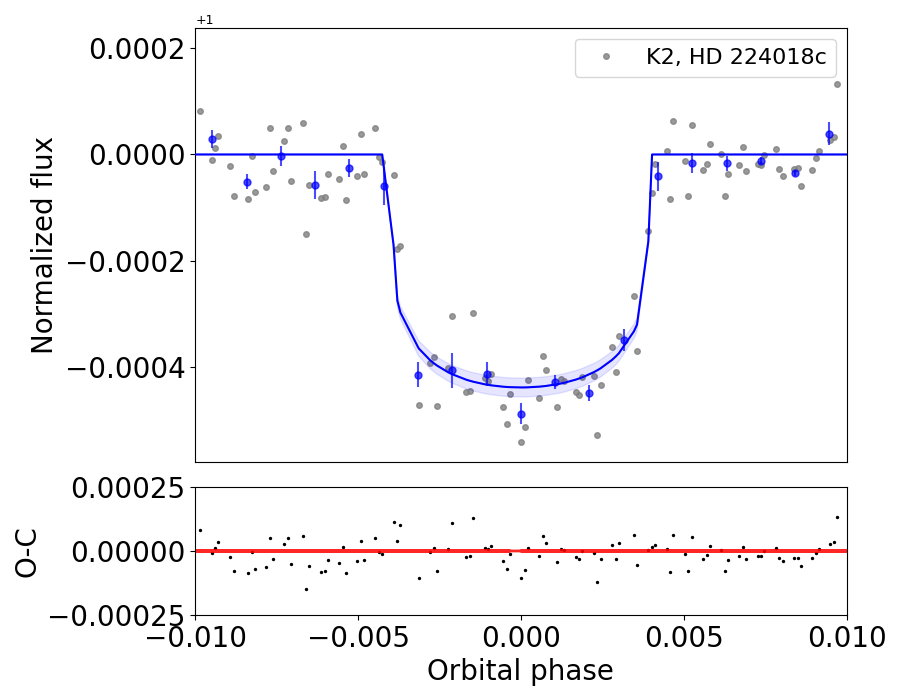}
        \includegraphics[width=0.33\textwidth]{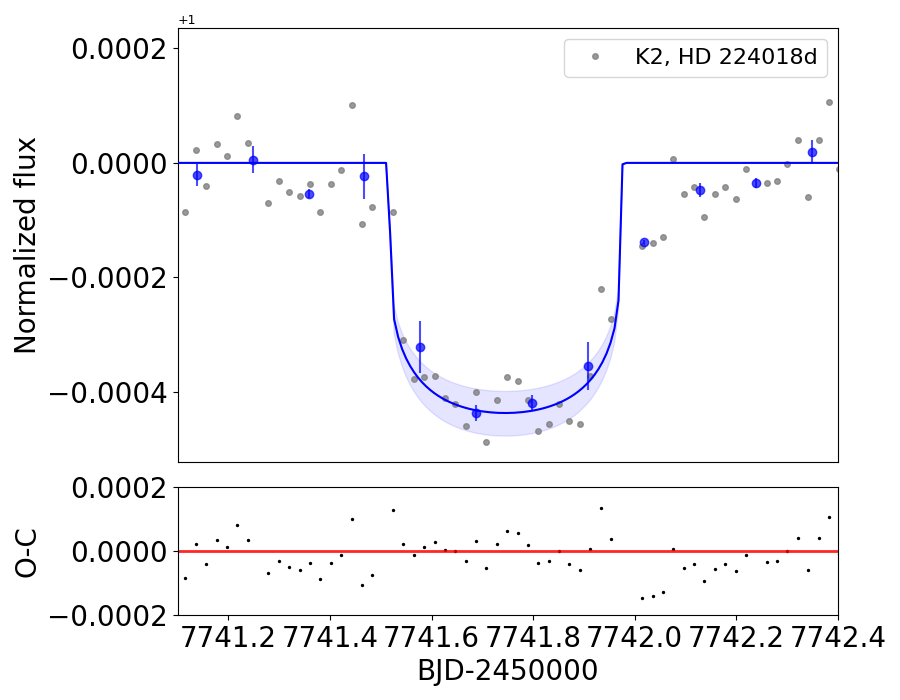}
    \caption{Transits of the three innermost planets in the HD\,224018 system as observed by K2. For planets b and c the transits are phase-folded to their corresponding orbital periods. The blue lines represent the best-fit transit models. The blue shaded area indicate the $1\sigma$ uncertainty in the transit depths (r$_p$/R$_\star)^2$. 
    The O-C data points are the residuals of the best-fit models.}
    \label{fig:transitmodel}
\end{figure*}

\begin{figure}[h!]
    \centering
    \includegraphics[trim={1cm 0 0.4cm 0},clip,width=0.45\textwidth]{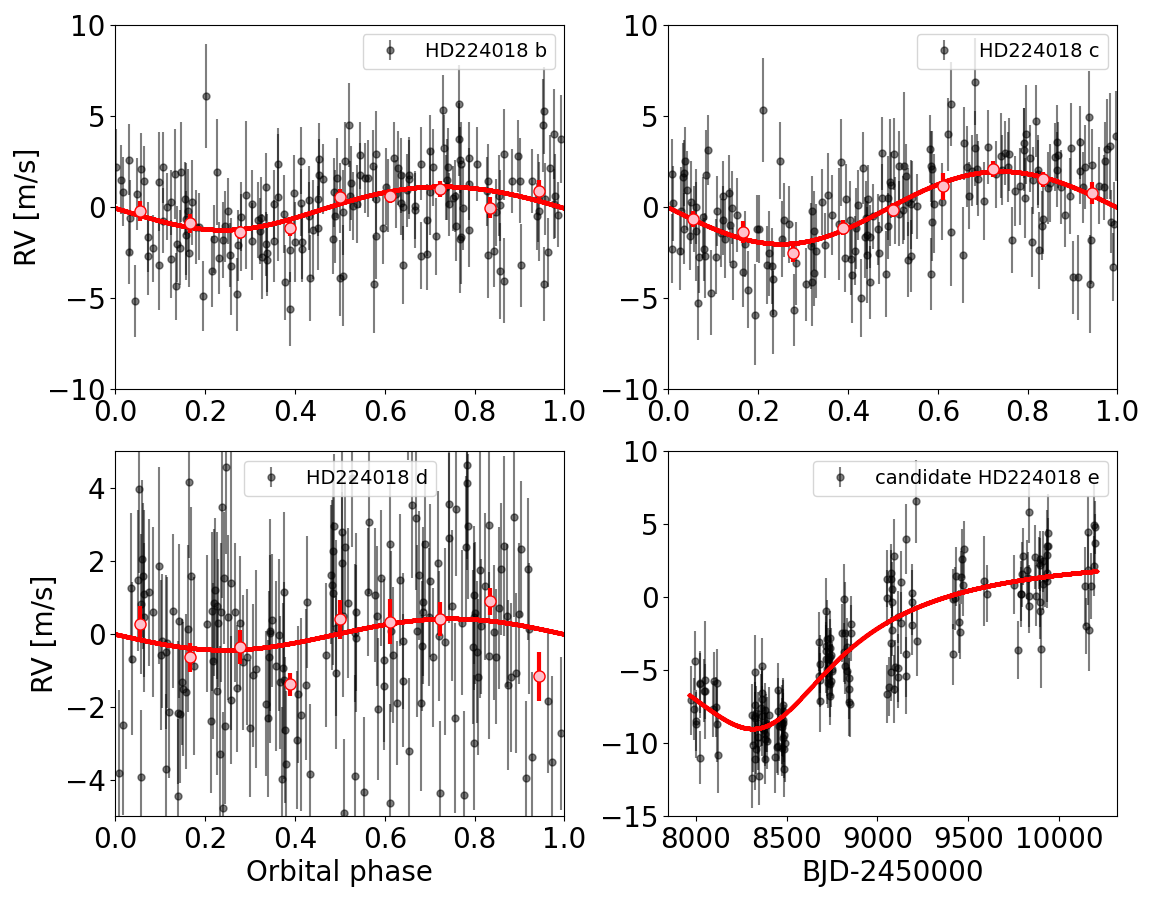}
    \caption{Doppler signals of HD\,224018 induced by the four planetary companions. For planets $b$, $c$, and $d$ phase folded data are shown. The best-fit models are indicated by a red curve. For HD\,224018\,d we used the maximum likelihood values of the Keplerian parameters ($K_d=0.43$ \ms; $P_d=138.07$ d; $T_c=2457741.745$ BJD; $e_d=0.018$; $\omega_{\rm \star,\,d}=2.74$ rad). For HD\,224018\,e the RV time series is shown because the orbit is not fully characterised. The error bars include the uncorrelated jitter added in quadrature to the RV uncertainties.} 
    \label{fig:rvplots}
\end{figure}

\begin{figure}
    \centering
    \includegraphics[trim={0 0 1cm 0},clip, width=0.5\textwidth]{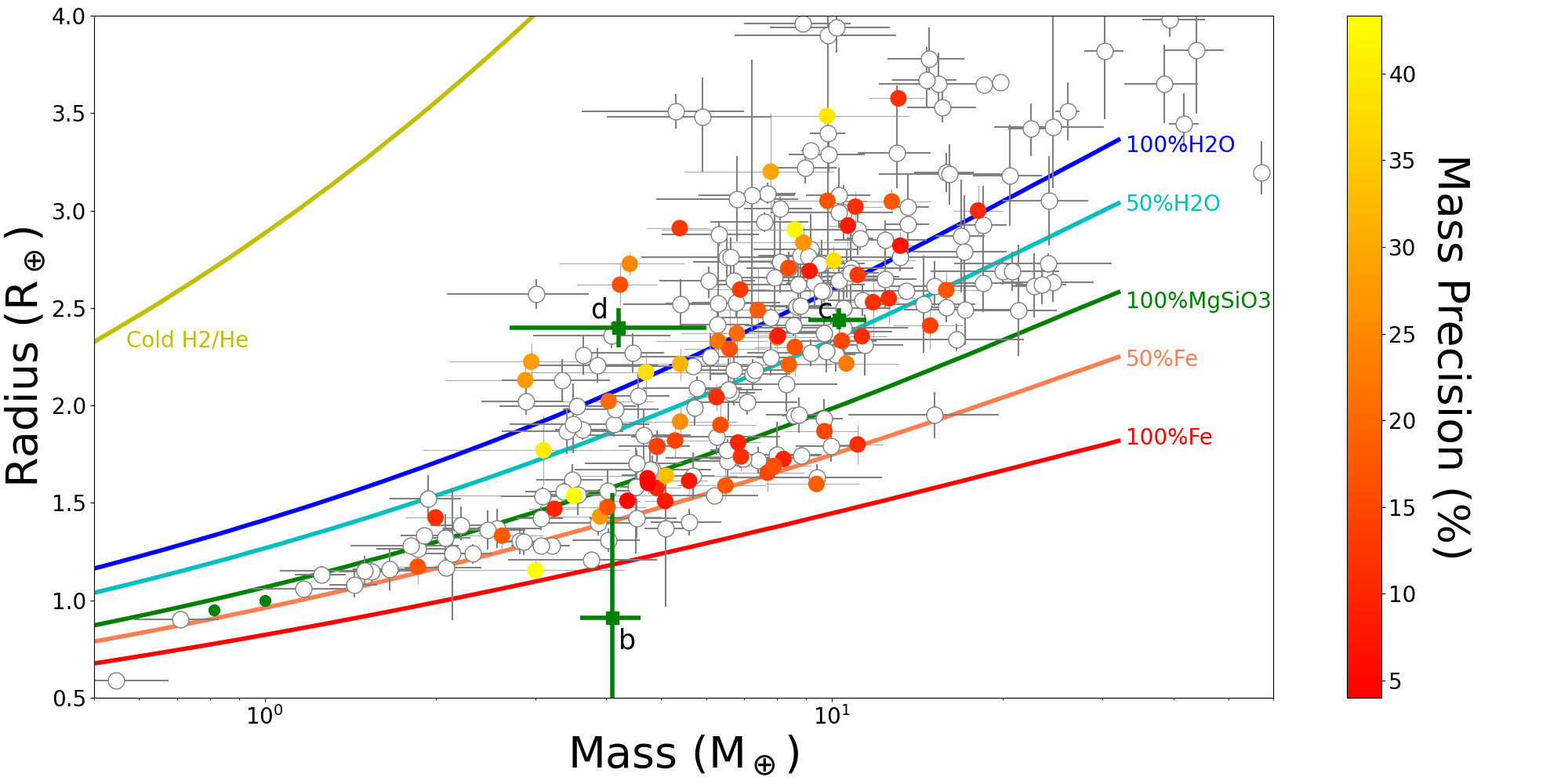}
    \caption{Mass-radius diagram for all planets with radii $R<4$\,\rearth\ characterised by HARPS-N (coloured dots, with the colour scale based on the precision of mass measurement). Non coloured dots correspond to all other planets with $R<4$\,\rearth\ and mass precision better than 30\% as measured by means of RVs. The three transiting planets orbiting HD~224018 are indicated with green squares. Green dots indicate the location of Earth and Venus. Some compositional theoretical curves are also plotted \citep{zeng2019PNAS..116.9723Z}. }
    \label{fig:mrdiagram}
\end{figure}

We show in Fig. \ref{fig:transitmodel} and \ref{fig:transitmodel2} the transit light curves of HD\,224018\,b, c, and d, and in Fig. \ref{fig:rvplots} the Doppler RV signals. The epochs of mid transit are summarised in Table \ref{tab:midtransittimes}. Fig. \ref{fig:rvplots2} shows the full RV time series and the four Keplerian model, after applying the \texttt{SCALPELS} correction. The cartoon in Fig. \ref{fig:tango} shows part of the K2 light curve at the epochs when the three innermost planets where transiting together. Fig. \ref{fig:mrdiagram} shows a mass-radius diagram including planets with precise mass measurements, the three transiting planets in the HD\,224018 system, and theoretical curves representative of some compositional models. 

\begin{table*}[ht]
\tiny
\centering
\caption{Priors and best-fit values for the free and derived parameters from the joint RV and light curve modeling.}
\label{tab:resultjointfit-compact}
\setlength{\tabcolsep}{1.2pt} % Riduce spazio tra colonne
\renewcommand{\arraystretch}{1.2} % Leggermente più spazio tra righe
\begin{tabularx}{\textwidth}{l | *{8}{>{\centering\arraybackslash}X}}
\toprule
\textbf{Parameter} & \multicolumn{2}{c}{\textbf{HD\,224018\,b}} & \multicolumn{2}{c}{\textbf{HD\,224018\,c}} & \multicolumn{2}{c}{\textbf{HD\,224018\,d}} & \multicolumn{2}{c}{\textbf{HD\,224018\,e}} \\
 & Prior & Best-fit value & Prior & Best-fit value & Prior & Best-fit value & Prior & Best-fit value\\
\midrule
\multicolumn{9}{c}{\textit{Free parameters}} \\
\midrule
$K$ [\ms] & $\mathcal{U}$(0,3) & 1.19$\pm$0.22 & $\mathcal{U}$(0,5) & $1.99^{+0.23}_{-0.20}$ & $\mathcal{U}$(0,5) & $0.52^{+0.22}_{-0.20}$ & $\mathcal{U}$(0,10) & $5.7^{+0.4}_{-0.3}$ \\
orbital period, $P$ [d] & $\mathcal{U}$(10.6,10.7) & 10.6413$\pm$0.0028 & $\mathcal{U}$(36.4,36.7) & $36.57669^{+0.00019}_{-0.00017}$ & $\mathcal{U}$(40,1000) & $138.0731^{+27.6127}_{-0.0050}$ & $\mathcal{U}$(1500,20000) & $9129^{+2499}_{-2479}$ \\
T$_{\rm conj\,or\,periastron}^{(a)}$ & $\mathcal{U}$(62.6,63.4) & $62.98^{+0.24}_{-0.22}$ & $\mathcal{U}$(78.0,78.5) & $78.1544^{+0.0080}_{-0.0084}$ & $\mathcal{U}$(41.4,42.1) & $41.745^{+0.051}_{-0.040}$ & $\mathcal{U}{\rm (-1700,14300)}$ & $692^{+67}_{-59}$ \\
eccentricity, $e$ & $\mathcal{N}$(0,0.1) & $0.06^{+0.07}_{-0.04}$ & $\mathcal{N}$(0,0.1) & 0.02$\pm$0.02 & $\mathcal{N}$(0,0.1) & $0.04^{+0.05}_{-0.04}$ & — & — \\
$\omega_\star$ [rad] & $\mathcal{U}$(0,2$\pi$) & $3.4^{+1.2}_{-1.5}$ & $\mathcal{U}$(0,2$\pi$) & $4.0^{+1.4}_{-1.9}$ & $\mathcal{U}$(0,2$\pi$) & $3.6^{+1.7}_{-2.4}$ & — &  — \\
$\sqrt{e}\cos\omega_\star$ & — & — & — & — & — & — & $\mathcal{U}$(-1,1) & $-0.743^{+0.059}_{-0.046}$ \\
$\sqrt{e}\sin\omega_\star$ & — & — & — & — & — & — & $\mathcal{U}$(-1,1) & -0.19$\pm$0.12 \\
$r/R_{\star}$ & $\mathcal{U}$(0,0.03) & 0.0073$^{+0.0051}_{-0.0046}$ & $\mathcal{U}$(0,0.03) & 0.01935$\pm$0.00039 & $\mathcal{U}$(0,0.03) & $0.01927^{+0.00087}_{-0.00085}$ & — & — \\
$i$ [deg] & $\mathcal{U}$(85, 90) & $86.7^{+1.6}_{-1.1}$ & $\mathcal{U}$(85, 90) & 89.8$\pm$0.2 & $\mathcal{U}$(85, 90) & 89.9$\pm$0.1 & — & — \\
\midrule
\multicolumn{9}{c}{\textit{Derived parameters}} \\
\midrule
radius, $r$ [\rearth] & — & 0.91$^{+0.64\, (+1.56)}_{-0.57\, (-0.90)}$\,$^{(b)}$ & — & $2.42^{+0.07}_{-0.08}$ & — & $2.4\pm0.1$ & — & — \\
mass, $m$ [\mearth] & — & $4.1\pm0.8$ & — & $10.4^{+1.3}_{-1.1}$ & — & $4.2^{+1.8}_{-1.6}$ (<9.0)$^{(b)}$ & — & $151^{+13}_{-14}$\,$^{(c)}$ \\
bulk density, $\rho$ [\gcm] & — & $17.1^{+10.2}_{-6.0}$ & — & $4.0^{+0.7}_{-0.6}$ & — & $1.7^{+0.8}_{-0.6}$ (<4.2)$^{(b)}$ & — & — \\
eccentricity, $e$ & — & — & — & — & — & — & — & $0.60^{+0.07}_{-0.08}$ \\
$\omega_{\star}$ [rad] & — & — & — & — & — & — & — & -2.9$^{+0.2}_{-0.1}$ \\
$a$ [au] & — & $0.0952^{+0.002}_{-0.002}$ & — & $0.217\pm0.005$ & — & $0.53^{+0.06}_{-0.02}$ & — & $8.6^{+1.5}_{-1.6}$ \\
$T_{\rm eq}$$^{(d)}$ [K] & — & $968\pm19$ & — & $641\pm12$ & — & $411^{+11}_{-21}$ & — & 102$^{+11}_{-8}$ \\
\midrule
\multicolumn{9}{c}{\textbf{Stellar and instrument-related parameters}} \\
\midrule
& \multicolumn{2}{c}{Prior} & \multicolumn{2}{c}{Best-fit value} \\
\midrule
$\rho_{\rm \star}$ $[\rm \rho_{\odot}]$ & \multicolumn{2}{c}{$\mathcal{N}(0.67, 0.07)$} & \multicolumn{2}{c}{0.64 $\pm$ 0.05} \\
$\alpha_{\rm 0}^{(e)}$ & \multicolumn{2}{c}{$\mathcal{U}(-5, 5)$} & \multicolumn{2}{c}{$0.13^{+0.07}_{-0.06}$} \\
$q_{\rm 1,\,K2\,,\,CHEOPS}$ & \multicolumn{2}{c}{$\mathcal{U}(0, 1)$} & \multicolumn{2}{c}{$0.30^{+0.19}_{-0.13}$} \\
$q_{\rm 2,\,K2\,,\,CHEOPS}$ & \multicolumn{2}{c}{$\mathcal{U}(0, 1)$} & \multicolumn{2}{c}{$0.28^{+0.27}_{-0.19}$} \\
$u_{\rm 1,\,K2\,,\,CHEOPS}$ & \multicolumn{2}{c}{derived} & \multicolumn{2}{c}{$0.31^{+0.21}_{-0.20}$} \\
$u_{\rm 2,\,K2\,,\,CHEOPS}$ & \multicolumn{2}{c}{derived} & \multicolumn{2}{c}{$0.23^{+0.26}_{-0.28}$} \\
$q_{\rm 1,\,TESS}$ & \multicolumn{2}{c}{$\mathcal{U}(0,1)$} & \multicolumn{2}{c}{$0.04^{+0.10}_{-0.03}$} \\
$q_{\rm 2,\,TESS}$ & \multicolumn{2}{c}{$\mathcal{U}(0,1)$} & \multicolumn{2}{c}{$0.27^{+0.32}_{-0.20}$} \\
$u_{\rm 1,\,TESS}$ & \multicolumn{2}{c}{derived} & \multicolumn{2}{c}{$0.10^{+0.14}_{-0.07}$} \\
$u_{\rm 2,\,TESS}$ & \multicolumn{2}{c}{derived} & \multicolumn{2}{c}{$0.08^{+0.17}_{-0.10}$} \\
$\gamma_{\rm K2}$ & \multicolumn{2}{c}{$\mathcal{U}(-10^{-3},10^{-3})$} & \multicolumn{2}{c}{-2.0$\pm$8.0$\cdot10^{-6}$} \\
$\gamma_{\rm CHEOPS}$  & \multicolumn{2}{c}{$\mathcal{U}(-10^{-3},10^{-3})$} & \multicolumn{2}{c}{-1.7$\pm$8.0$\cdot10^{-5}$}  \\
$\gamma_{\rm TESS}$  & \multicolumn{2}{c}{$\mathcal{U}(-10^{-3},10^{-3})$} & \multicolumn{2}{c}{-4$\pm$4$\cdot 10^{-5}$}  \\
$\sigma_{\rm jit,\,K2}$  & \multicolumn{2}{c}{$\mathcal{U}(0,10^{-3})$} & \multicolumn{2}{c}{6$\pm$4$\cdot10^{-6}$}  \\
$\sigma_{\rm jit,\,CHEOPS}$  & \multicolumn{2}{c}{$\mathcal{U}(0,10^{-3})$} & \multicolumn{2}{c}{1.66$\pm$8.0$\cdot10^{-5}$}  \\
$\sigma_{\rm jit,\,TESS}$  & \multicolumn{2}{c}{$\mathcal{U}(0,10^{-3})$} & \multicolumn{2}{c}{2.93$\pm$6.0$\cdot10^{-4}$}  \\
$\sigma_{\rm jit,\,HARPS\,N}$ [ms$^{-1}$] & \multicolumn{2}{c}{$\mathcal{U}(0,5)$]} & \multicolumn{2}{c}{1.4$\pm$0.1}  \\
$\gamma_{\rm HARPS\,N}$ [ms$^{-1}$] & \multicolumn{2}{c}{$\mathcal{U}(-5,5)$} & \multicolumn{2}{c}{4.0$^{+0.7}_{-0.8}$}  \\
\bottomrule
\end{tabularx}

\tablefoot{\tiny
				\tablefoottext{a}{Epochs are given as BJD-2457700. For planets b, c, and d they correspond to the times of inferior conjunction. For the candidate planet e, this epoch is the time of periastron passage.}
                \tablefoottext{b}{Values in parentheses are the uncertainties and upper limits at the 99.7\% confidence level.}
                \tablefoottext{c}{This is the minimum mass $m_e \sin i_e$ [\mearth].}
                 \tablefoottext{d}{Assuming zero albedo. For planet e, the semi-major axis is taken as a reference distance for the calculation.}
                \tablefoottext{e}{This is the \texttt{SCALPELS} coefficient.}
			}
\end{table*}

\section{Internal structure analysis of HD~224018\,c} \label{sec:internalstructure}

As a next step, we used the \texttt{plaNETic}\footnote{\url{https://github.com/joannegger/plaNETic}} framework \citep{Egger+2024} to infer the internal structure of HD~224018\,c. We focused only on this planet because it is the one for which we could measure a precise mass and radius. We did run models also for planets~b~and~d, but given the low precision in radius and mass the inferred posteriors are very close to the chosen priors. This effect is even stronger for planet~d, as here also the orbital period and therefore the planet's equilibrium temperature is not well constrained. \texttt{plaNETic} uses a neural network trained on the forward model of \texttt{BICEPS} \citep{Haldemann+2024} as a surrogate model in combination with a full grid accept-reject sampling scheme, allowing for a fast and reliable characterisation of the internal structure of observed exoplanets.

For this purpose, the planet is modelled as a combination of an inner core (Fe, S), a mantle (oxidised Si, Mg and Fe) and a volatile layer (uniformly mixed H/He and water). As the connection between the composition of planets and their host stars is still highly debated in the literature \citep[e.g.][]{Thiabaud+2015,Adibekyan+2021,Michel+2020}, we ran models using three different compositional priors, one assuming the planetary Si/Mg/Fe ratios equal the ones of the host star \citep{Thiabaud+2015}, one assuming the planet is iron-enriched compared to the host star \citep{Adibekyan+2021}, and one independent of the stellar abundances, where the planetary Si/Mg/Fe ratios are sampled using a uniform prior. Additionally, we also used two different priors for the water content of the planet, resulting in a water-rich scenario as expected if the planet would have formed outside the ice-line, and a water-poor scenario in accordance with a formation inside the ice-line. More details on framework and chosen priors can be found in \cite{Egger+2024}.

\begin{figure*}
    \centering
    \includegraphics[width=0.8\linewidth]{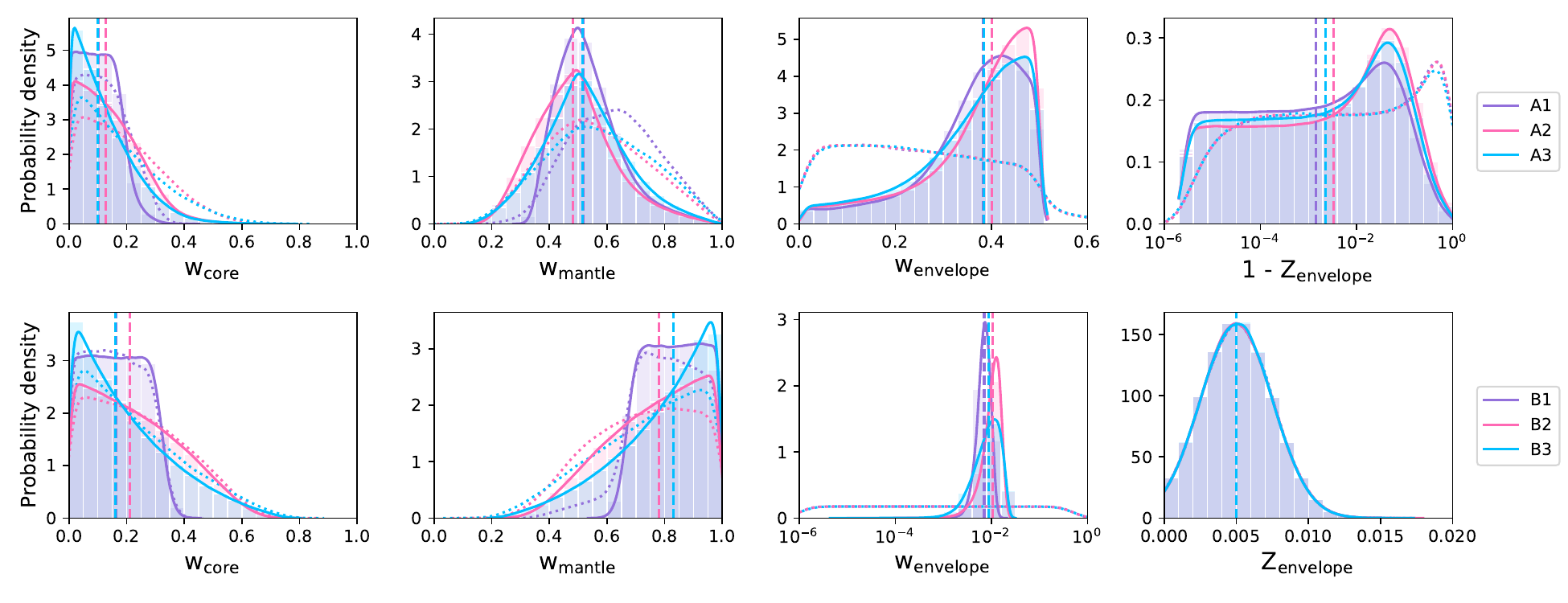}
    \caption{Posterior distributions for internal structure parameters of HD~224018~c: mass fractions, with respect to the total planet mass, of i) the inner core (w$_{\rm core}$); ii) the mantle (w$_{\rm mantle}$); iii) the envelope layers (w$_{\rm envelope}$); mass fraction of H/He in the envelope (1-Z$_{\rm envelope}$); mass fraction of water in the envelope layer (Z$_{\rm envelope}$). Dotted lines show the priors, dashed vertical lines the median values of the posteriors. Different colours are used for different compositional priors for the planetary Si/Mg/Fe ratios, which were chosen to be stellar (purple), iron-enriched compared to the host star (pink) or freely sampled using a uniform prior (blue). The top row shows models generated using a water prior in agreement with a formation outside the ice-line, the bottom row with a formation inside the ice-line.}
    \label{fig:int_struct_c}
\end{figure*}

Figure~\ref{fig:int_struct_c} shows the most important internal structure parameters for HD~224018~c. If we assume the planet formed in an environment where water could only be accreted through the accreted gas (i.e. inside the water ice-line), we infer a rather tightly constrained envelope of around 1\% in total planet mass that is almost purely made up of H/He. On the other hand, for a water-prior consistent with a formation scenario outside the ice-line we find that a wide range of envelope mass fractions is possible, with a preference for more massive envelopes. Generally, these envelopes show high water mass fractions.

\begin{table*}[h!]
\renewcommand{\arraystretch}{1.4}
\tiny
\caption{Results of the internal structure modeling for HD~224018~c.}
\centering
\begin{tabular}{r|ccc|ccc}
\hline \hline
Water prior &              \multicolumn{3}{c|}{Water-rich (formation outside ice-line)} & \multicolumn{3}{c}{Water-poor (formation inside ice-line)} \\
Si/Mg/Fe prior &           Stellar (A1) &       Iron-enriched (A2) &      Free (A3) &
                           Stellar (B1) &       Iron-enriched (B2) &      Free (B3) \\
\hline
w$_\textrm{core}$ [\%] &        $10_{-7}^{+7}$ &    $13_{-9}^{+12}$ &    $10_{-7}^{+12}$ &
                           $16_{-11}^{+11}$ &    $21_{-15}^{+20}$ &    $16_{-12}^{+21}$ \\
w$_\textrm{mantle}$ [\%] &      $52_{-9}^{+12}$ &    $48_{-13}^{+14}$ &    $52_{-13}^{+15}$ &
                           $83_{-11}^{+11}$ &    $78_{-20}^{+15}$ &    $83_{-21}^{+12}$ \\
w$_\textrm{envelope}$ [\%] &    $38_{-12}^{+8}$ &    $40_{-15}^{+7}$ &    $38_{-15}^{+8}$ &
                           $0.7_{-0.2}^{+0.2}$ &    $1.1_{-0.4}^{+0.4}$ &    $0.9_{-0.5}^{+0.6}$ \\
\hline
Z$_\textrm{envelope}$ [\%] &        $99.9_{-4.8}^{+0.1}$ &    $99.7_{-7.3}^{+0.3}$ &    $99.8_{-6.2}^{+0.2}$ &
                           $0.5_{-0.2}^{+0.2}$ &    $0.5_{-0.2}^{+0.2}$ &    $0.5_{-0.2}^{+0.2}$ \\
\hline
x$_\textrm{Fe,core}$ [\%] &     $90.3_{-6.4}^{+6.5}$ &    $90.4_{-6.4}^{+6.5}$ &    $90.3_{-6.4}^{+6.5}$ &
                           $90.3_{-6.4}^{+6.5}$ &    $90.4_{-6.4}^{+6.5}$ &    $90.4_{-6.4}^{+6.5}$ \\
x$_\textrm{S,core}$ [\%] &      $9.7_{-6.5}^{+6.4}$ &    $9.6_{-6.5}^{+6.4}$ &    $9.7_{-6.5}^{+6.4}$ &
                           $9.7_{-6.5}^{+6.4}$ &    $9.6_{-6.5}^{+6.4}$ &    $9.6_{-6.5}^{+6.4}$ \\
\hline
x$_\textrm{Si,mantle}$ [\%] &   $42_{-6}^{+7}$ &    $38_{-10}^{+9}$ &    $37_{-26}^{+29}$ &
                           $42_{-6}^{+7}$ &    $38_{-10}^{+10}$ &    $36_{-25}^{+29}$ \\
x$_\textrm{Mg,mantle}$ [\%] &   $41_{-6}^{+7}$ &    $37_{-9}^{+9}$ &    $37_{-26}^{+32}$ &
                           $41_{-6}^{+7}$ &    $36_{-10}^{+10}$ &    $36_{-25}^{+30}$ \\
x$_\textrm{Fe,mantle}$ [\%] &   $17_{-11}^{+9}$ &    $24_{-16}^{+19}$ &    $17_{-13}^{+22}$ &
                           $17_{-11}^{+9}$ &    $25_{-17}^{+19}$ &    $19_{-14}^{+24}$ \\
\hline
\end{tabular}
\label{tab:internal_structure_results_c}
\end{table*}
\renewcommand{\arraystretch}{1.0}

\section{Summary and conclusions} \label{sec:conclusions}

We presented the discovery and first characterisation of a multi-planetary system orbiting the 7.0$^{+3.4}_{-3.2}$ Gyr old Sun-analogue HD\,224018. The system includes three planet-size transiting companions, observed to transit simultaneously in one circumstance (Fig. \ref{fig:tango}). A fourth outermost family member is detected only in the RVs. The innermost planet $b$ ($P_b$=10.6413$\pm$0.0028 d), was initially detected from the analysis of the RVs alone, and its transits were then revealed in the K2 data, although they are shallow and characterised by a low S/N, preventing us from an accurate and precise modeling. Our analysis reveals that it has a mass $m_{\rm b}$=4.1$\pm$0.8 \mearth, and a radius $r_{\rm b}$=0.91$^{+0.64\,(+1.56)}_{-0.57\,(-0.90)}$ \rearth\, (3$\sigma$ error bars are given in parenthesis).  
The planet's location on the mass-radius diagram indicates that the planet's internal composition spans from that of the Earth down to a bare 100$\%$ iron core. 

The second planet in order of distance from the host star is a warm sub-Neptune, and it is the only member of the system for which we can provide a precise mass and radius (relative uncertainties of $1.6\%$ and 2.1$\%$, respectively), with a bulk density of 3.9$\pm$0.5 \gcm. That allowed us to model the interior structure of HD\,224018\,c, with the results that depend on whether the planet formed inside or outside the water ice-line. 

The third companion, HD\,224018\,d, was first discovered thanks to a mono-transit observed in the K2 light curve. We found that there is a likely second transit in TESS data, and our analysis shows that the most likely orbital period is $\sim$138 days. While this paper was under review, a new TESS light curve was issued (Sector 92), which will be analysed in detail in a future study. Only a transit of planet c is detected, while a transit event possibly imputable to planet d is not seen. This non-detection is in agreement with the prediction assuming an orbital period of 138.07 days. The new TESS observations seem to rule out an orbital period of 146.19 days, which is the value immediately following 138.07 days in the grid of possible periods compatible with the transits observed by K2 and TESS (see Fig. \ref{fig:plbsingletrans}). The planet is a sub-Neptune with a well measured radius, $r_d$=2.4$\pm0.1$ \rearth, and we derived the dynamical mass $m_d$=4.2$^{+1.8}_{-1.6}$ \mearth\, ($<9$ \mearth\, at $99.7\%$ significance level). 
The analysis of the available data shows that HD\,224018\,d has a very similar size to HD\,224018\,c, but the precision on the mass measurement do not allow us to constrain its internal structure and composition. 

For the fourth and outermost member of the system, the current dataset does not allow us to well constrain its wide orbit, although this appears to be highly eccentric ($e_e=0.60^{+0.07}_{-0.08}$). We present HD\,224018\,e as a candidate giant companion with a minimum mass nearly half that of Jupiter, a semi-major axis 8.6$^{+1.5}_{-1.6}$~au, and the periastron at $\sim 3.5$~au. The Gaia DR3 Renormalised Unit Weight Error (RUWE) of HD~224018 is 0.92 (single-star solution). The sensitivity curves to companions on wide orbits (Fig. \ref{fig:ruwe}) show that massive Jupiters at a$\sim$2--3 au, and brown dwarfs at a$>$5 au can be ruled out, and that a planet with mass $\sim$0.5 M$_{\rm Jup}$ is well below the detectability threshold.
A RV follow-up is required to establish if HD\,224018\,e belongs to the population of cold ($a\sim1-10$~au) or ultra-cold ($a \gtrsim 10$~au) Jupiters.  
In the former case, HD\,224018 will add to the current sample of only fifteen systems with transiting inner low-mass planets ($m<$20 \mearth) and outer cold Jupiters \citep{Bonomo2025A&A...700A.126B}; it may also be one of the very few systems with multiple inner small planets in the presence of a high eccentricity ($e \gtrsim 0.4$) cold Jupiter. 

\begin{acknowledgements}
The HARPS-N project was funded by the Prodex Program of the Swiss Space Office (SSO), the Harvard University Origin of Life Initiative (HUOLI), the Scottish Universities Physics Alliance (SUPA), the University of Geneva, the Smithsonian Astrophysical Observatory (SAO), the Italian National Astrophysical Institute (INAF), University of St. Andrews, Queen’s University Belfast, and University of Edinburgh.
We used data from the European Space Agency (ESA) mission {\it Gaia} (\url{https://www.cosmos.esa.int/gaia}), processed by the {\it Gaia} Data Processing and Analysis Consortium (DPAC, \url{https://www.cosmos.esa.int/web/gaia/dpac/consortium}). Funding for the DPAC has been provided by national institutions, in particular the institutions participating in the {\it Gaia} Multilateral Agreement. \textit{Kepler} was competitively selected as the 10th Discovery mission. Funding for the \textit{Kepler}/K2 missions were provided by NASA’s Science Mission Directorate.
CHEOPS is an ESA mission in partnership with Switzerland with contributions to the payload and the ground segment from Austria, Belgium, France, Germany, Hungary, Italy, Portugal, Spain, Sweden and the UK.
This paper includes data collected by the TESS mission. Funding for the TESS mission is provided by the NASA Explorer Program. We acknowledge the use of public TOI Release data from pipelines at the TESS Science Office and at the TESS Science Processing Operations Center. Resources supporting this work were provided by the NASA High-End Computing (HEC) Program through the NASA Advanced Supercomputing (NAS) Division at Ames Research Center for the production of the SPOC data products.
We used The Data $\&$ Analysis Center for Exoplanets (DACE), which is a facility based at the University of Geneva (CH) dedicated to extrasolar planets data visualisation, exchange and analysis. DACE is a platform of the Swiss National Centre of Competence in Research (NCCR) PlanetS, federating the Swiss expertise in Exoplanet research. The DACE platform is available at \url{https://dace.unige.ch}. 
A.M. acknowledges funding from a UKRI Future Leader Fellowship, grant number MR/X033244/1 and a UK Science and Technology Facilities Council (STFC) small grant ST/Y002334/1. A.C.C acknowledges support from STFC consolidated grant number ST/V000861/1 and UKRI/ERC Synergy Grant EP/Z000181/1 (REVEAL). L.M. acknowledges support from the European Union- NextGenerationEU (PRIN MUR 2022 20229R43BH). R.T. acknowledges the Head of Department of Physics, the University of Cambridge, and the Carlsberg Foundation for supporting this research. C.A.W. would like to acknowledge support from the UK Science and Technology Facilities Council (STFC, grant number ST/X00094X/1).
We acknowledge support from the INAF Large Grant 2023 ``EXODEMO'', and from the Swiss National Science Foundation within the framework of the NCCR PlanetS under grants 51NF40\_182901 and 51NF40\_205606. 
\end{acknowledgements}

\bibliographystyle{aa} % style aa.bst
\bibliography{aa55770-25} % your references Yourfile.bib

\appendix

\section{YARARA applied to HARPS-N spectra} \label{app:yarara}
The pipeline begins from the S1D order-merged spectra level produced by the official DRS v3.0.1 \citep{dumusque2021A&A...648A.103D}, with a minimal requirement of prior information during the processing in order to allow an easy scalability with different instruments. 
The extracted information contains as much as activity indicators, than stellar atmospheric parameters \citep{cretignier2024MNRAS.535.2562C}, even if the final goal of the method is to provide an independent measurement of the RV time-series, either measured by a CCF using a tailored line-selection \citep{cretignier2020A&A...633A..76C} or line-by-line RVs \citep{cretignier2023A&A...678A...2C} following the prescription of \citep{dumusque2018A&A...620A..47D}. 
The analysis with \texttt{YARARA} revealed that an offset visible in the CCF area time series around BJD 2\,458\,500 is mainly due to the blue part of the spectra, a result similar to the one obtained on the solar telescope (see the orange curve in Fig. A1 of \citealt{klein2024MNRAS.531.4238K}) and likely related to an instrumental systematic on the detector. The RVs of HD\,224018 appear to be affected by an offset around the same epoch, which is spotted by the analysis performed with SCALPELS discussed in Sect. \ref{sec:yararascalpels} (see Fig. \ref{fig:rvtimeseries}). We can exclude that the overall long-term variability seen in the RV time series, at a level of a few \ms, is determined by systematics related to the long-term stability of the HARPS-N spectrograph, like instrumental RV drift, instrument warm-up, or power failure, which have been found to produce offsets no larger than 2 $\ms$ \citep{meunier2024A&A...687A.303M}. This leads us to conclude that the long-term trend observed in RVs has mainly an astrophysical origin, such as the presence of a long-term substellar companion.

\section{Additional plots and tables}

\begin{figure}[h!]
    \centering
    \includegraphics[width=0.5\textwidth]{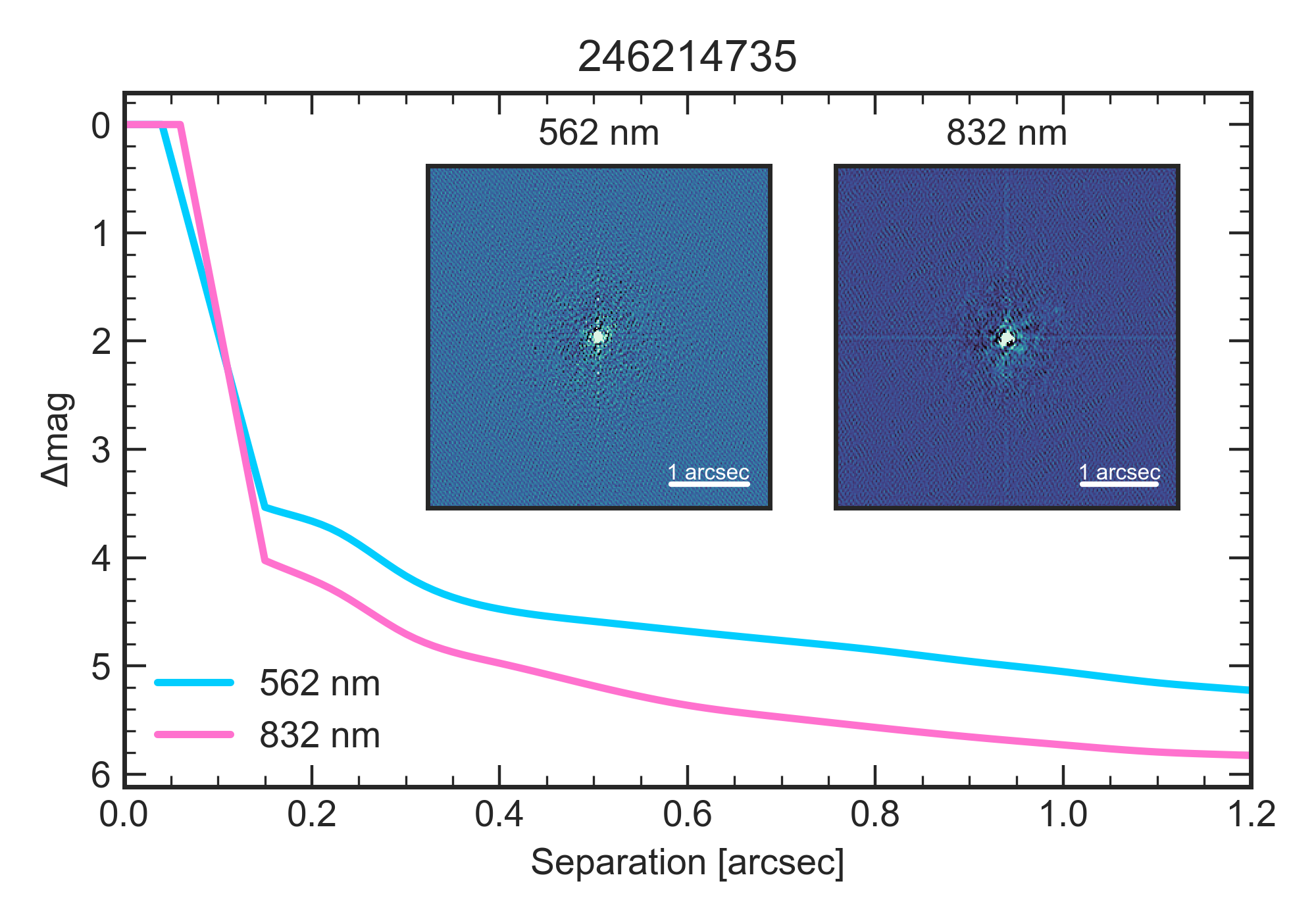}
    \caption{HD\,224018 observed with the NESSI speckle imager mounted on the 3.5\,m WIYN telescope at Kitt Peak.}
    \label{fig:wiyn}
\end{figure}

\begin{figure}[h!]
\centering
\includegraphics[width=1\linewidth]{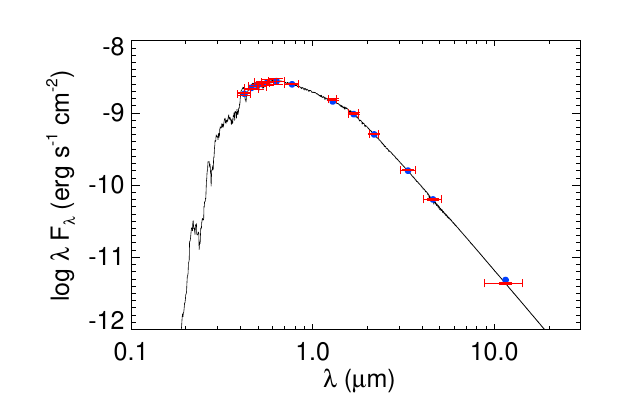}
\caption{Stellar spectral energy distribution. The broad band measurements from the Tycho, APASS Johnson, Sloan, 2MASS and WISE magnitudes are shown in red, and the corresponding theoretical values with blue circles. The best-fit model is displayed with a black solid line.} 
\label{fig:stellarSED}
\end{figure}

\begin{figure*}
    \centering
    \includegraphics[width=0.9\textwidth]{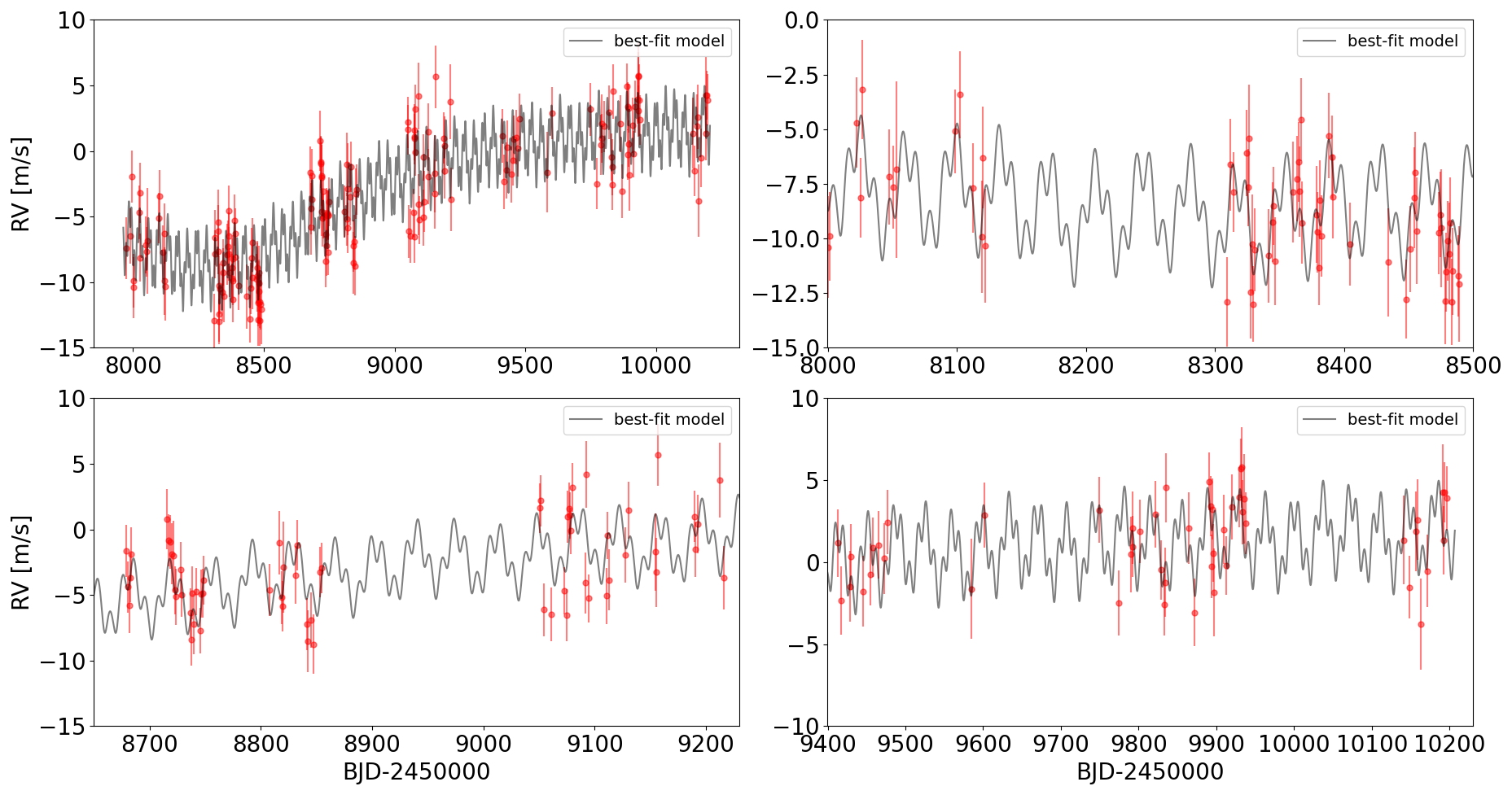}
    \caption{Time series of the RVs. The measurements are indicated by red dots, and our best-fit spectroscopic model is over-plotted with a black line. The upper left panel shows the full time series, the other plots show zoomed-in views to better appreciate the agreement between the observations and best-fit model. The error bars include the uncorrelated jitter added in quadrature to the RV uncertainties.} 
    \label{fig:rvplots2}
\end{figure*}

\begin{figure}
    \centering
    \includegraphics[width=0.5\textwidth]{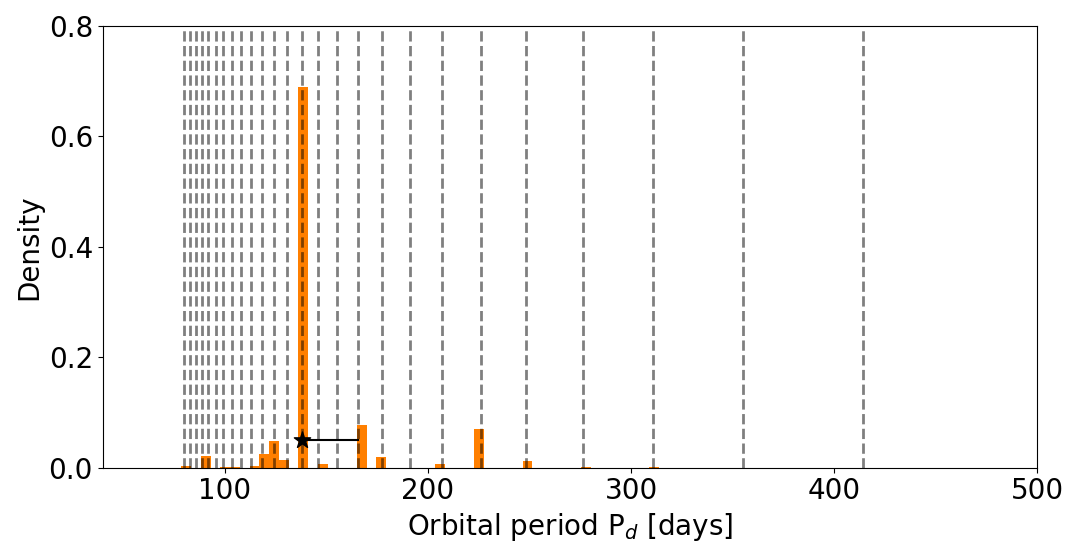}
    \caption{Posterior of the orbital period $P_d$ obtained from a joint RV and photometry modeling. The width of each bin is $\sim$5 days. The best-fit value ($\pm1\sigma$) is indicated with a black asterisk. Dashed vertical lines correspond to possible orbital periods from a grid of discrete $P_d$ values calculated from the transit midpoints from K2 and TESS.}
    \label{fig:posterior_pd}
\end{figure}

\begin{figure}
        \includegraphics[width=0.5\textwidth]{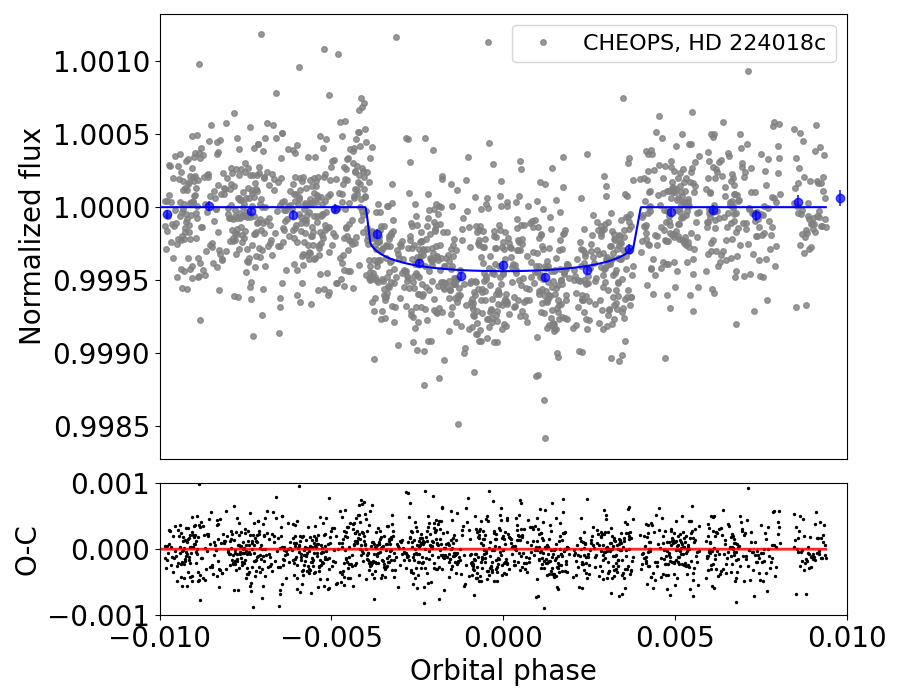}\\
        \includegraphics[width=0.5\textwidth]{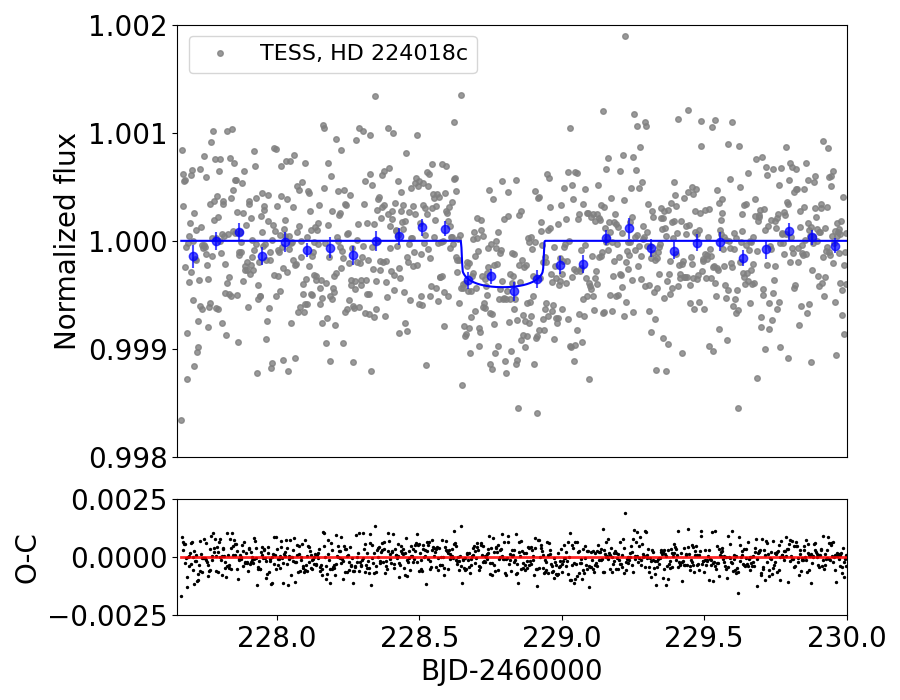}\\
       \includegraphics[width=.5\textwidth]{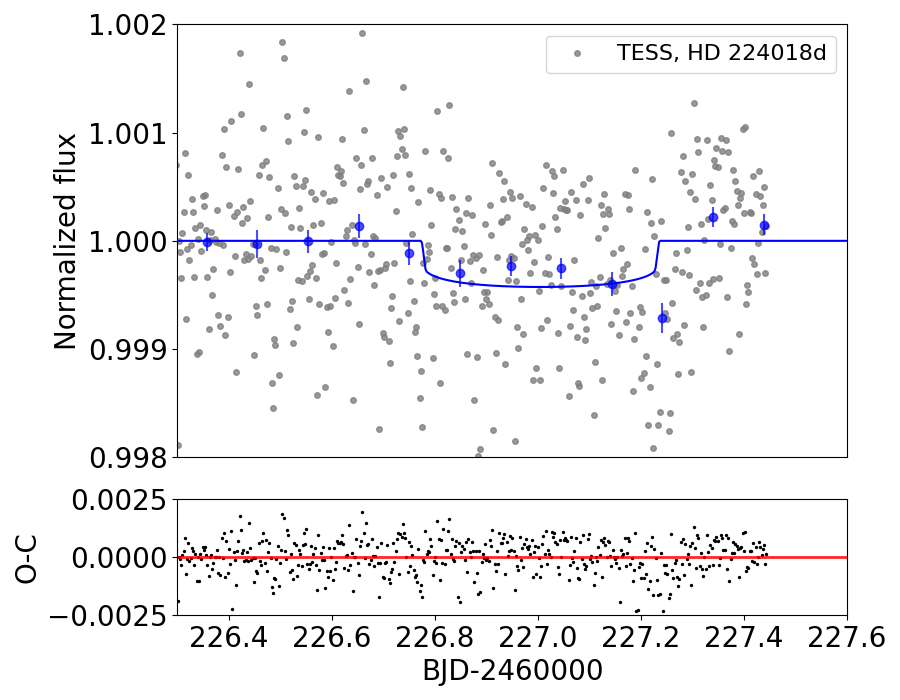}
    \caption{Transits of HD\,224018\,c in CHEOPS, and TESS photometry (first and second panels). Transit of HD\,224018\,d observed by TESS (lower panel). 
    Residuals (O-C) after subtracting the best-fit transit models are shown for each transit.}
    \label{fig:transitmodel2}
\end{figure}

\begin{figure*}
    \centering
    \includegraphics[width=\textwidth]{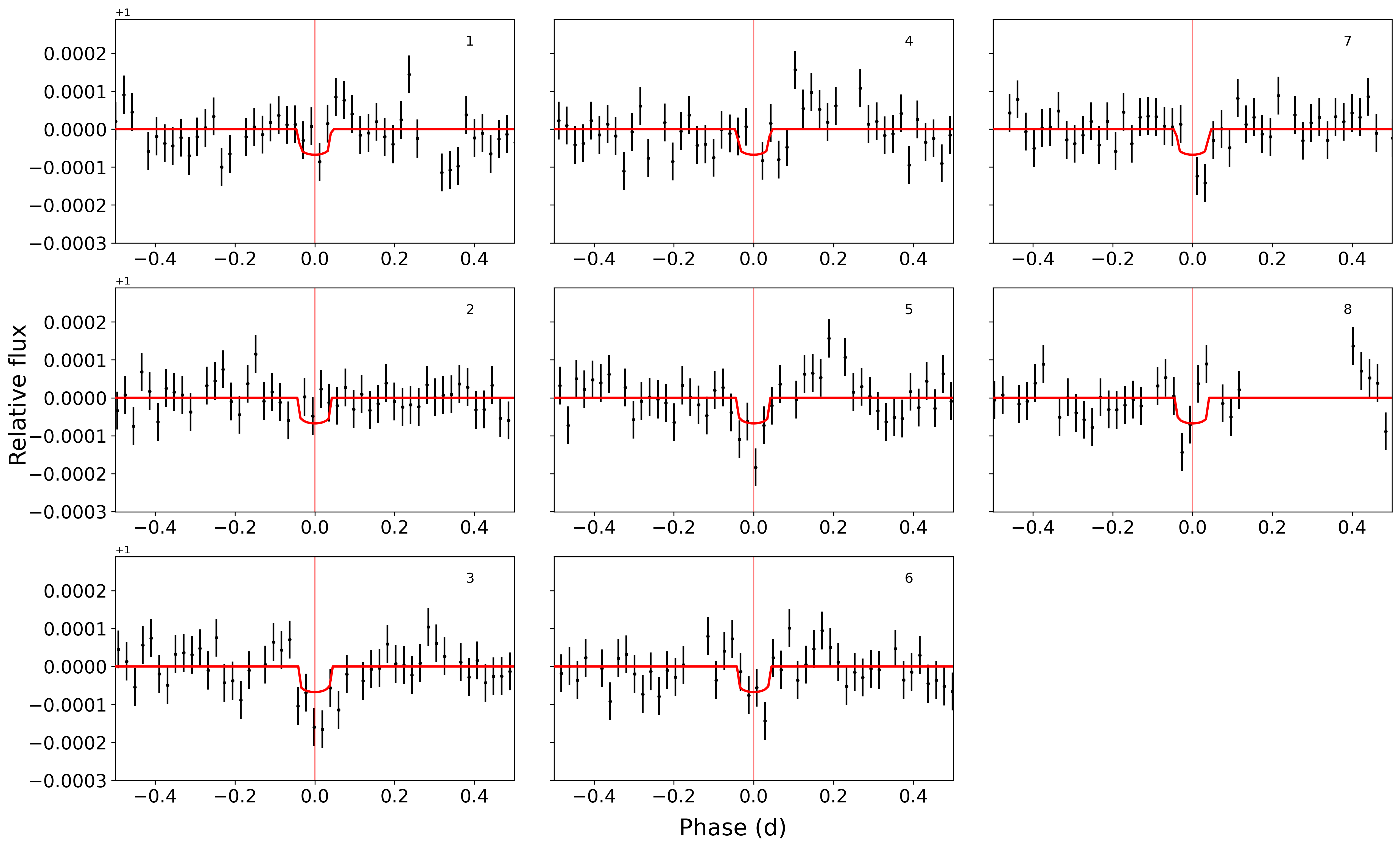}
    \caption{Individual transits of HD\,224018\,b in the K2 light curve, following the order specified in the top right label of each panel. The best-model fit is represented by the red line. Note that the light curve has been cleared of both the systematics and the transits of the other two planets.}
    \label{fig:plbsingletrans}
\end{figure*}

\begin{figure}
\centering
\includegraphics[width=0.45\textwidth]{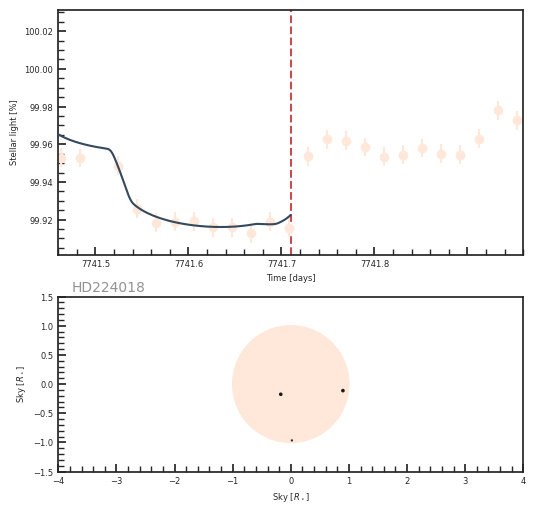}
\caption{Snapshot of the triple transit of planet b, c, and d as seen by K2 at the end of 2016. This artistic representation has been created with the software \texttt{tango} (https://github.com/oscaribv/tango). }
\label{fig:tango}
\end{figure}

\begin{figure}
    \centering
    \includegraphics[width=0.5\textwidth]{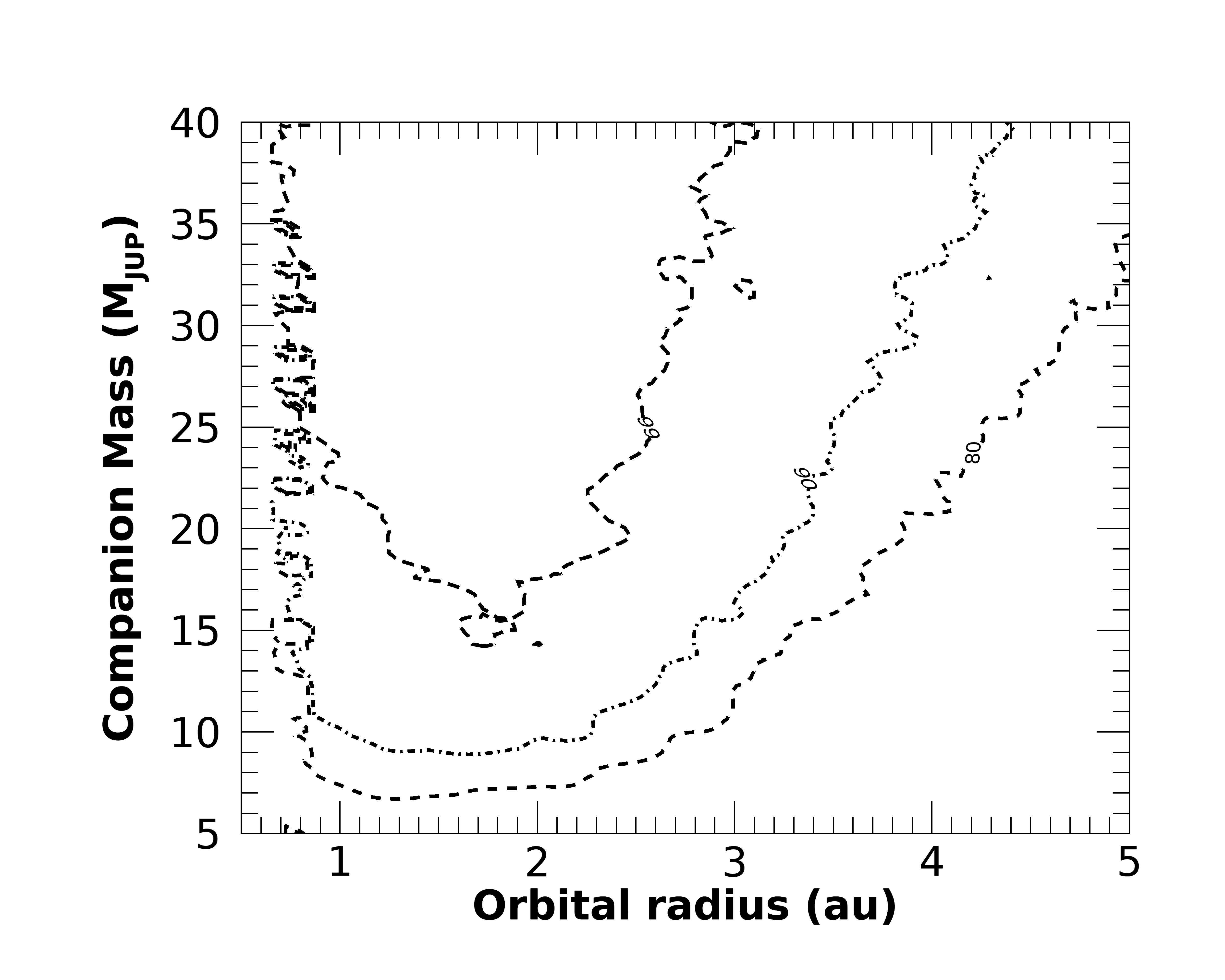}
    \caption{Gaia DR3 sensitivity to companions of a given mass as a function of the orbital semi-major axis orbiting HD 224018. Dashed, dashed-dotted, and long-dashed lines correspond to iso-probability curves for 80, 90, and 99\% probability of a companion of given properties to produce RUWE$>$0.92.}
    \label{fig:ruwe}
\end{figure}

\begin{table}[]
    \centering
    \begin{tabular}{ccc}
    \hline
    \noalign{\smallskip}
    Planet  & Time of mid transit & Telescope\\
    &  (BJD-2\,450\,000) & \\
    \noalign{\smallskip}
    \hline
    \noalign{\smallskip}
     HD\,224018\,b  & $7741.701^{+0.170}_{-0.018}$ & K2\\
      \noalign{\smallskip}
     & $7752.308^{+0.147}_{0.027}$ & K2\\
 \noalign{\smallskip}
     & $7762.988^{+0.018}_{-0.013}$ & K2 \\
      \noalign{\smallskip}
     & $7773.620^{+0.013}_{-0.015}$ & K2 \\
      \noalign{\smallskip}
     & $7784.257^{+0.013}_{-0.011}$ & K2 \\
      \noalign{\smallskip}
     & $7794.917^{+0.012}_{-0.011}$ & K2 \\
      \noalign{\smallskip}
     & $7805.569^{+0.0170}_{-0.013}$ & K2 \\
      \noalign{\smallskip}
     & $7816.168^{+0.042}_{-0.020}$ & K2 \\
      \noalign{\smallskip}
     \hline
      \noalign{\smallskip}
     HD\,224018\,c  & 7741.5812$^{+0.0020}_{-0.0019}$ & K2\\
      \noalign{\smallskip}
     & 7778.1547$^{+0.0027}_{-0.0031}$ & K2 \\
      \noalign{\smallskip}
     & 7814.7357$^{+0.0035}_{-0.0026}$ & K2 \\
      \noalign{\smallskip}
     & 9460.6843$^{+0.0026}_{-0.0024}$ & CHEOPS \\
      \noalign{\smallskip}
     & 9497.2615$^{+0.0021}_{-0.0024}$ & CHEOPS \\
      \noalign{\smallskip}
     & 10228.7957$^{+0.0065}_{-0.0068}$ & TESS \\
      \noalign{\smallskip}
      \noalign{\smallskip}
     \hline
      \noalign{\smallskip}
     HD\,224018\,d & 7741.7465$^{+0.0039}_{-0.0037}$ & K2 \\
      \noalign{\smallskip}
     & 10227.0583$^{+0.0039}_{-0.0037}$ & TESS \\
     \noalign{\smallskip}
    \hline
    \end{tabular}
    \caption{Epochs of mid transits for the three innermost planets in the system. Values are calculated after fitting each transit individually.}
    \label{tab:midtransittimes}
\end{table}

\end{document}